\documentclass[a4paper,12pt]{article}

\def\theequation{\arabic{section}.\arabic{equation}}
\newcounter{rown}
\def\bl{\setcounter{rown}{\value{equation}}
\stepcounter{rown}\setcounter{equation}0
\def\theequation{\thesection.\arabic{rown}\alph{equation}}}
\def\el{\setcounter{equation}{\value{rown}}
\def\theequation{\thesection.\arabic{equation}}}

\usepackage[centertags]{amsmath}
\usepackage{amsfonts}
\usepackage{amssymb}
\usepackage{amsthm}
\usepackage{newlfont}
\usepackage{appendix}
\def\bZ{{\bar Z}}
\def\dalpha{{\dot\alpha}}
\def\dbeta{{\dot\beta}}

\def\opi{{\overline{\pi}}}

\def\of{{\overline{f}}}

\def\adb{{\alpha\dbeta}}

\begin{document}
\begin{flushright}
FTUV--05--1028 \quad IFIC--05--56 \ \  IFT UWr 0210/05
\\[1cm]
\end{flushright}

\begin{center}
\vspace{.7cm} {\Large {\bf Extension of the Shirafuji model for
Massive Particles with Spin }}

\bigskip

\renewcommand{\thefootnote}{\alph{footnote}}

{\bf Sergey Fedoruk}${}^{+}$\footnote{E-mail: {\tt
fed@postmaster.co.uk}}, {\bf Andrzej
Frydryszak}${}^{\dag}$\footnote{E-mail: {\tt
amfry@ift.uni.wroc.pl}}, {\bf Jerzy Lukierski}
${}^{\dag}$\footnote{E-mail:
{\tt lukier@ift.uni.wroc.pl}},\\
{\bf C\`{e}sar
Miquel-Espanya}${}^{\dag}$\,${}^{\ddag}$\footnote{E-mail: {\tt
Cesar.Miquel@ific.uv.es}}

\vskip 0.4truecm

\vskip 0.2cm ${}^{+}$\ {\it Ukrainian Engineering-Pedagogical
Academy, Kharkov, Ukraine}

\vskip 0.2cm ${}^{\dag}$\ {\it Institute of Theoretical Physics,
Wroc{\l}aw University, 50-204 Wroc{\l}aw, Poland }

\vskip 0.2cm ${}^{\ddag}$\ {\it Departamento de F\'{\i}sica
Te\'orica and IFIC (Centro Mixto CSIC-UVEG), 46100-Burjassot
(Valencia), Spain}
\end{center}

\begin{abstract}
We extend the Shirafuji model for massless particles with primary
spacetime coordinates and composite four-momenta to a model for
massive particles with spin and electric charge. The primary
variables in the model are the spacetime four-vector, four scalars
describing spin and charge degrees of freedom as well as a pair of
Weyl spinors. The geometric description proposed in this paper
provides an intermediate step between the free purely twistorial
model in two-twistor space in which both spacetime and four-momenta
vectors are composite, and the standard particle model, where both
spacetime and four-momenta vectors are elementary. We quantize the
model and find explicitly the first-quantized wavefunctions
describing relativistic particles with mass, spin and electric
charge. The spacetime coordinates in the model are not commutative;
this leads to a wavefunction that depends only on one covariant
projection of the spacetime four-vector (covariantized time
coordinate) defining  plane wave solutions.
\end{abstract}

\section{Introduction}

\setcounter{footnote}{0}

There are known three  equivalent ways of describing
massless relativistic particles:

i) \textit{Purely twistorial description} - with primary twistor
 variables and composite both spacetime and four-momenta. A free point particle
moving in twistor space
$Z^A=(\omega^\alpha,\opi_\dbeta)\in\mathbb{C}^4$ ($A=1,\dots,4$;
$\alpha,\dbeta=1,2$) (see e.g. \cite{Hu79}) is described by the
action
\begin{equation}\label{TwistorialModel}
S_1=\frac{i}{2}\int d\tau\left[\left(
\bar{Z}^A\dot{Z}_A-h.c.\right)+\lambda\left(\bar{Z}^A
Z_A\right)\right]\, ,
\end{equation}
where $\bZ_A$ denotes the complex conjugation of $Z^A$, the
conformal $SU(2,2)$ scalar product is $\bar{Z}^A {Z}_A \equiv
\bar{Z}_A g^{AB} Z_B$  and the conformal metric $g^{AB}$ in twistor
space is usually chosen to be $g^{AB}=\left(
 \begin{array}{cc}
o & -iI_2 \\
iI_2 & 0 \\
\end{array}
\right)$.

ii) \textit{Mixed twistorial-spacetime description} - with primary
spacetime coordinates and composite four-momenta.  The relativistic
phase space \sloppy $(x^{\adb},P_\adb)=(\frac{1}{2}(\sigma_\mu)^\adb
x^\mu,\frac{1}{2}(\sigma_\mu)^\adb P^\mu)$ is determined by the
basic relations of the  Penrose theory\footnote{We use the following
notation. The metric is mostly minus
$\eta_{\mu\nu}={\text{diag}}(+---)$. The Weyl two-spinor indices are
risen and lowered in the following way
$\varphi^\alpha=\epsilon^{\alpha\beta}\varphi_\beta$,
$\varphi_\alpha=\varphi^\beta\epsilon_{\beta\alpha}$,
$\bar\varphi^{\dot\alpha}=\epsilon^{\dot\alpha\dot\beta}\bar\varphi_{\dot\beta}$,
$\bar\varphi_{\dot\alpha}=\bar\varphi^{\dot\beta}\epsilon_{\dot\beta\dot\alpha}$
where
$\epsilon^{\alpha\beta}\epsilon_{\beta\gamma}=-\delta^{\alpha}_{\gamma}$,
$\epsilon^{\dot\alpha\dot\beta}\epsilon_{\dot\beta\dot\gamma}=
-\delta^{\dot\alpha}_{\dot\gamma}$. The algebra for the
$\sigma$-matrices
$\sigma^\mu_{\alpha\dot\beta}=(\overline{\sigma^\mu_{\beta\dot\alpha}})$
and $\sigma_\mu^{\dot\alpha\alpha}=\epsilon^{\alpha\beta}
\epsilon^{\dot\alpha\dot\beta}\sigma_{\mu\beta\dot\beta}$ is
$\sigma_{\mu\alpha\dot\gamma}\sigma_\nu^{\dot\gamma\beta}+
\sigma_{\nu\alpha\dot\gamma}\sigma_\mu^{\dot\gamma\beta}=
2\eta_{\mu\nu}\delta_{\alpha}^{\beta}$;
$\sigma_{\mu\alpha\dot\alpha}\sigma^\mu_{\beta\dot\beta}=
2\epsilon_{\alpha\beta}\epsilon_{\dot\alpha\dot\beta}$. Also we
define $A_\mu=\sigma_{\mu\alpha\dot\beta}A^{\alpha\dot\beta}$, and
therefore
$A_{\alpha\dot\beta}=\frac{1}{2}A_{\mu}\sigma^\mu_{\alpha\dot\beta}$,
$A^{\dot\alpha\beta}=\frac{1}{2}A^{\mu}\sigma_\mu^{\dot\alpha\beta}$
for any vector $A_{\mu}$.} \bl
\begin{eqnarray}
P_{\adb}&=&\pi_\alpha\pi_\dbeta\, ,\label{Padb-1twistor}\\
\omega^\alpha&=&ix^{\adb}\pi_\dbeta\, ,\label{PenroseIncidence}
\end{eqnarray}
\el where the constraint $Z^A \bZ_A=0$ is required if we wish
$x^{\adb}$ to be Hermitian ({\it i.e.} $x^\mu=(\sigma^\mu)_\adb
x^\adb$ is real). In the mixed twistor-spacetime approach we use
only the relation (\ref{PenroseIncidence}), and we obtain from
(\ref{TwistorialModel}) (modulo divergence term) the Shirafuji model
\cite{Sh83} $(\dot{a}\equiv \frac{da}{dt})$
\begin{equation}\label{ShirafugiModel}
S'_1=\int d\tau \pi_\alpha\opi_\dbeta \dot{x}^{\adb}\, ,
\end{equation}
which was extensively used by the Kharkov group (see e.g.
\cite{So89}).
\\
iii) \textit{Standard geometric description} - with primary relativistic
 phase space variables (spacetime coordinates and four-momenta). Inserting in
(\ref{ShirafugiModel}) the relation (\ref{Padb-1twistor}) we obtain
the known action for the massless relativistic particle moving in
Minkowski space
\begin{equation}\label{StandardModel}
S''_1=\int d\tau \left(P_\adb \dot{x}^\adb -eP^2\right)\, ,
\end{equation}
 where the  term $eP^2$ encodes the constraint
$P^2 = 2P_\adb P^\adb=0$ following algebraically from
(\ref{Padb-1twistor}). If we eliminate $P_\mu$, we obtain the
standard Lagrangian for a free massive relativistic particle
\begin{equation}\label{StandardModel-b}
S'''_1 = \frac{1}{2}\, \int d\tau \, \frac{1}{e} \, \dot{x}_\mu \,
 \dot{x}^\mu \, .
\end{equation}

The three equivalent models (\ref{TwistorialModel}),
(\ref{ShirafugiModel}) and (\ref{StandardModel}) describe three
possible different geometric set-ups in the theory of massless
relativistic particles: a purely spinorial (twistorial) framework, a
hybrid formulation using simultaneously spinorial and spacetime
elementary coordinates, and the description in the  standard
relativistic phase space $(x_\mu,P_\mu)$.

The extension of these three geometric levels to the two-twistor
sector has been presented recently
\cite{BeAzLuMi04,BeLuMi04,AzFrLuMi} in terms of the corresponding
Liouville one-forms. If we introduce two twistors ($i=1,2$;
$A=1,\dots,4$)\footnote{The indices $i=1,2$ describe an internal
$SU(2)$ symmetry. The complex conjugation implies the change from
covariant (lower) indices to contravariant (upper) indices.}
\begin{equation}
Z_{A i}= (\omega^\alpha_{\phantom{\alpha}i},\bar\pi_{\dalpha
i})\, ,
\end{equation}
the free Liouville one-form extending (modulo   constraints) the
action (\ref{TwistorialModel}) to the two-twistor case is the
following
\begin{equation}
\Theta_2=\frac{i}{2}\left(\omega^\alpha_{\phantom{\alpha}i}d\pi_{\alpha
i}+ \bar\pi_{\dot{\alpha}i}d\bar\omega^{\dalpha i}-h.c.\right)\, .
\end{equation}
After using the two-twistor generalization of (\ref{Padb-1twistor}),
(\ref{PenroseIncidence}) \bl
\begin{eqnarray}
P_\adb&=&\pi_{\alpha}^{i}\bar\pi_{\dbeta i}\quad
,\label{Padb-2twistor}\\
\omega^\alpha_{\phantom{\alpha}i}&=& iz^{\adb}\bar\pi_{\dbeta
i}\, ,\label{PenroseIncidence.2twistor}
\end{eqnarray}
\el where
\begin{equation}
z^{\adb}=x^\adb + i y^\adb\label{Complex.Space-Time}\, ,
\end{equation}
one obtains
\begin{equation}
\Theta'_2=\pi_\alpha^{\phantom{\alpha}i}\bar\pi_{\dbeta i}dx^\adb+
iy^\adb(\pi_\alpha^{\phantom{\alpha}i}d\bar\pi_{\dbeta i}-
\bar\pi_{\dbeta i} d\pi_\alpha^{\phantom{\alpha}i})\, . \label{LiouvillePrime2}
\end{equation}
We introduce the new variables
\begin{equation}
s^i_{\ j}=-2y^\adb \pi_{\alpha}^{i}\bar\pi_{\dot\beta
j}=\overline{(s^j_{\ i})}\, ,\label{sij}
\end{equation}
and define $f$ and $\of$ satisfying \bl
\begin{eqnarray}
\bar\pi^{\dalpha
i}\bar\pi_\dalpha^{\phantom{\alpha}j}&=&-\epsilon^{ij}f\quad
,\label{pipi1}\\
\pi^\alpha_{\phantom{\alpha}i}\pi_{\alpha j}&=&-\epsilon_{ij}\bar
f\quad
,\label{pipi2}\\
\bar\pi_{\dalpha
i}\bar\pi_{\dbeta}^{\phantom{\alpha}i}&=&\epsilon_{\dalpha\dbeta}f\quad
,\label{pipi3}\\
\pi_{\alpha i}\pi_\beta^{\phantom{\beta} i}
&=&\epsilon_{\alpha\beta}\bar f\, .\label{pipi4}
\end{eqnarray}
\el Inverting (\ref{sij})
\begin{equation}
y^\adb=-\frac{1}{2|f|^2}s_i^{\ j}\pi^{\alpha
i}\opi^{\beta}_{\phantom{\beta}j}\, ,
\end{equation}
and using (\ref{Padb-2twistor}) one obtains from
(\ref{LiouvillePrime2})
\begin{equation}\label{LiouvillePrime2New}
\Theta'_2=P_\mu dx^\mu+\frac{i}{2}s_i^{\ j}\left[\frac{1}{\bar
f}\pi^\alpha_{\phantom{\alpha}k}d\pi_{\alpha j}\epsilon^{ki}
+\frac{1}{f}\opi^{\dalpha
i}d\bar\pi_{\dalpha}^{\phantom{\alpha}k}\epsilon_{kj}\right]\, .
\end{equation}

The formula (\ref{LiouvillePrime2New}) determines the two-twistor
generalization of the Shirafuji action (\ref{ShirafugiModel}). We
see that the primary, or equivalently elementary, variables are now
the following ones
\begin{eqnarray}
N=1 &\Rightarrow& N=2\nonumber\\
x^\adb, \pi_\alpha,\bar\pi_\dalpha&\Rightarrow& x^\adb,\pi_{\alpha
i},\bar\pi_\dalpha^{\phantom{\alpha}i},s_i^{\ j}\, .
\end{eqnarray}
The particle model described by the Liouville one-form
(\ref{LiouvillePrime2New}) provides a framework to describe the
mass, spin and electric charge but does not specify their values. We
shall introduce further their numerical values by postulating
suitable physical constraints. One concludes that the
quantum-mechanical solution of the model (\ref{LiouvillePrime2New})
may describe infinite-dimensional higher spin and electric charge
multiplets, linked with the field-theoretic formulation of higher
spin theories (see e.g. \cite{Vas,Sez,Sor}).

The plan of our paper is the following:

In Sect. \ref{Sect.An.Constaints} we define our model by its
kinematic part following from (\ref{LiouvillePrime2New}) and by
adding four physical constraints. We shall describe the model in the
corresponding phase space with the enlarged spacetime sector $(Q_L;
L=1,\dots,16)$ and the enlarged momenta $(\mathcal{P}_L;
L=1,\dots,16)$\footnote{The subindex $A$ enumerates the real degrees
of freedom.}\bl
\begin{eqnarray}
Q_L&=&(x^\adb,\pi_{\alpha i},\bar\pi^{\dalpha i},s_i^{\ j})\quad
,\label{Phase.Space.Coord}\\
\mathcal{P}_L&=&(P_\adb,P^{\alpha i},\bar{P}_{\dalpha
i},P_{(s)}{}^{i}{}_{j})\, .\label{Phase.Space.Momenta}
\end{eqnarray}
\el
We also present the complete classical analysis of the
constraints.

In Sect. \ref{Sect.Sol.2nd.Class.Constr} we show how to eliminate
all the second class constraints by a  non linear change of the
variables $(x^\adb,P_\adb,P_{\alpha
i},\bar{P}_\dalpha^{\phantom{\alpha}i})$, {\it i.e.}, by choosing a
set of suitable coordinates in the phase space
(\ref{Phase.Space.Coord}), (\ref{Phase.Space.Momenta}), and by
introducing Dirac brackets.

In Sect. \ref{Sect.Sol.1st.Class.Constr} we introduce the first
quantization of the model and  provide the solution of the first
class constraints. In such a way we obtain the wave equations for
the massive particles with spin and electric charge. It appears that
the wavefunction is determined in the four-momentum space, because
in twistor formalism the composite four-momenta are commuting. The
composite spacetime coordinates $x_\mu$, defined by
(\ref{PenroseIncidence.2twistor}) and (\ref{Complex.Space-Time}),
are non-commutative due to the following Poisson bracket \cite{Be84}
\begin{equation}\label{X.Non-commutative}
\{x_\mu,x_\nu\}=\frac{1}{m^4}\epsilon_{\mu\nu\rho\tau}P_\rho
W_\tau\, ,
\end{equation}
where $W_\tau$ is the Pauli-Luba\'{n}ski four-vector. It follows,
however, from (\ref{X.Non-commutative}) that
\begin{equation}
\{P^\mu x_\mu,x_\nu\}=0\, ,
\end{equation}
what implies that  our quantum-mechanical wavefunction for the
non-vanishing spin case can depend on the projection
$\widetilde{\tau}=\frac{P^\mu x_\mu}{m}$ of the spacetime
four-vector defining covariantized scalar time coordinate.

In Sect. \ref{Sect.Examples} we present a brief outlook.

\section{The classical model - analysis of constraints in phase space}\label{Sect.An.Constaints}
\setcounter{equation}{0}

\subsection{Action, conservation laws and physical constraints}
We describe the dynamics of a massive spinning particle by its
trajectory in the generalized coordinate space
\begin{equation}\label{coord}
Q_L (\tau) =\left( x^\mu(\tau), \pi_{\alpha k}(\tau),
\bar\pi_{\dot\alpha}^k(\tau), s_k{}^j(\tau)\right)\, ,
\end{equation}
where $x^\mu$ is the spacetime vector of position, $\pi_{\alpha k}$,
$\bar\pi_{\dot\alpha}^k=\overline{(\pi_{\alpha k})}$ ($k,j=1,2$) are
two pairs of commuting Weyl spinors and the four quantities
$s_k{}^j$, satisfying the reality condition
$s_k{}^j=\overline{(s_j{}^k)}$, are Lorentz scalars. The action
derived from (\ref{LiouvillePrime2New}) has the following form
$(a=1,\dots,4)$
\begin{equation}\label{act}
S=\int d\tau\,{\cal L} = \int d\tau\,\left[P_\mu \dot x^\mu +
{\frac{i}{2}}s_k{}^j \left({\frac{1}{\bar f}}\pi^{\alpha
k}\dot\pi_{\alpha j}
+{\frac{1}{f}}\bar\pi^{\dot\alpha}_j\dot{\bar\pi}_{\dot\alpha}^k\right)
+\lambda^a T_a\right]\, ,
\end{equation}
where the $T_a$ are algebraic constraints on the coordinates
(\ref{coord}) to be specified later, and the quantities
\begin{equation}\label{P(pi)}
P^{\alpha\dot\alpha}=\pi^{\alpha k}\bar\pi^{\dot\alpha}_k\quad ;
\end{equation}
\begin{equation}\label{f}
f={\frac{1}{2}}\bar\pi_{\dot\alpha k}\bar\pi^{\dot\alpha k}\, ,
\qquad \bar f={\frac{1}{2}}\pi_{\alpha k}\pi^{\alpha k}\, ,
\end{equation}
are bilinear functions of the spinors $\pi_{\alpha k}$ and
$\bar\pi_{\dot\alpha}^k$.

{}From the Lagrangian (\ref{act}) we obtain\bl
\begin{eqnarray}\label{p-x}
P_\mu=\frac{\partial{\cal L}}{\partial\dot x^\mu}&=&
\sigma_{\mu\alpha\dot\beta}\pi^{\alpha
k}\bar\pi^{\dot\alpha}_k\, ,\\
\label{p-pi} P^{\alpha j}=\frac{\partial{\cal
L}}{\partial\dot\pi_{\alpha j}}= \frac{i}{2\bar f}\pi^{\alpha
k}s_k{}^j\, ,& & \bar P^{\dot\alpha}_j= \frac{\partial{\cal
L}}{\partial\dot{\bar\pi}_{\dot\alpha}^j} = \frac{i}{2f} s_j{}^k
\bar\pi^{\dot\alpha}_k \, ,\\
\label{p-s} P_{(s)}{}^k{}_j&=& \frac{\partial{\cal L}}{\partial\dot
s_k{}^j }= 0\, .
\end{eqnarray}\el

The formulae (\ref{p-x})-(\ref{p-s}), defining the  momenta, give us
the following sixteen {\it primary constraints}\bl
\begin{equation}\label{c:p-x}
P_\mu -\sigma_{\mu\alpha\dot\beta}\pi^{\alpha
k}\bar\pi^{\dot\alpha}_k\approx 0\, ,
\end{equation}
\begin{equation}\label{c:p-pi}
P^{\alpha j} -\frac{i}{2\bar f}\pi^{\alpha k}s_k{}^j \approx 0\quad
, \qquad \bar P^{\dot\alpha}_j - \frac{i}{2f} s_j{}^k
\bar\pi^{\dot\alpha}_k \approx 0\, ,
\end{equation}
\begin{equation}\label{c:p-s}
P_{(s)}{}^k{}_j \approx 0\, .
\end{equation}\el

In order to determine the variables which describe the spin degrees
of freedom we calculate the Noether charges $M_{\mu\nu}$
corresponding to the Lorentz symmetries for the action (\ref{act}).
We obtain
\begin{equation}\label{1noe}
M_{\mu\nu}=x_\mu P_\nu- x_\nu P_\mu-{\textstyle\frac{1}{2}}
P^{\alpha k}(\sigma_{\mu\nu})_\alpha{}^\beta \pi_{\beta k}+
{\textstyle\frac{1}{2}}\bar\pi_{\dot\alpha}^k
(\bar\sigma^{\mu\nu})^{\dot\alpha}{}_{\dot\beta} \bar
P^{\dot\beta}_k\, ,
\end{equation}
where $P_\mu$, $P^{\alpha k}$, $P^{\dot\alpha}_k$ are taken from
(\ref{p-x})-(\ref{p-s}) and the definitions for $\sigma^{\mu\nu}$
and $\overline{\sigma}^{\mu\nu}$ are
\begin{equation}\label{1bisnoe}
(\sigma^{\mu\nu})_\alpha{}^\beta\equiv {\textstyle\frac{1}{2}}
(\sigma^\mu_{\alpha\dot\gamma}\sigma^{\nu\dot\gamma\beta}-
\sigma^\nu_{\alpha\dot\gamma}\sigma^{\mu\dot\gamma\beta})\, ,
\qquad (\bar\sigma^{\mu\nu})^{\dot\alpha}{}_{\dot\beta}\equiv
{\textstyle\frac{1}{2}}( \sigma^{\mu\dot\alpha\gamma}
\sigma^\nu_{\gamma\dot\beta}- \sigma^{\nu\dot\alpha\gamma}
\sigma^\mu_{\gamma\dot\beta})\, .
\end{equation}
The Pauli-Luba\'{n}ski four-vector is given by the formula
\begin{equation}\label{1bbisnoe}
W^\mu={\textstyle\frac{1}{2}}\epsilon^{\mu\nu\lambda\rho} P_\nu
M_{\lambda\rho}={\textstyle\frac{i}{2}} P^{\alpha
k}(\sigma^{\mu\nu})_\alpha{}^\beta \pi_{\beta k} P_\nu
+{\textstyle\frac{i}{2}} \bar\pi_{\dot\alpha}^k
(\bar\sigma^{\mu\nu})^{\dot\alpha}{}_{\dot\beta} \bar
P^{\dot\beta}_k P_\nu\, ,
\end{equation}
where the relations
$\epsilon^{\mu\nu\lambda\rho}\sigma_{\lambda\rho}=-2i\sigma^{\mu\nu}$,
$\epsilon^{\mu\nu\lambda\rho}\bar\sigma_{\lambda\rho}=2i\bar\sigma^{\mu\nu}$
have been used. Inserting the expressions (\ref{p-x})-(\ref{p-s}) in
(\ref{1bbisnoe}) we obtain
\begin{equation}\label{2noe}
W^\mu = -{\textstyle\frac{1}{2}}P^\mu{}_k{}^l (s_l{}^k
+\epsilon^{kj}s_j{}^n \epsilon_{nl})\, ,
\end{equation}
where
\begin{equation}\label{3noe}
P^\mu{}_k{}^l \equiv \pi^\alpha_k
 \sigma^\mu_{\alpha\dot\beta}
\bar\pi^{\dot\beta l}\, .
\end{equation}
In fact, only the traceless part of $s_l{}^k$ is present in
(\ref{2noe}). So, if we insert in (\ref{2noe}) the decomposition
$s_l{}^k=s_0 \delta_l{}^k + s_r (\tau_r)_l{}^k$ we obtain
\begin{equation}\label{4noe}
W^\mu = -P^\mu{}_k{}^l (\tau_r)_l{}^k s_r \, .
\end{equation}
Now, using
\begin{equation}\label{4bisnoe}
P^\mu{}_k{}^l P_\mu{}_i{}^j=2\epsilon_{ki}\epsilon^{lj} f\bar f\quad
,
\end{equation}
we obtain
\begin{equation}\label{5noe}
W^\mu W_\mu = -4 f\bar f {\mathbf{s}}^2 \, .
\end{equation}
But from (\ref{P(pi)})-(\ref{f})  we can derive
\begin{equation}\label{5bisnoe}
4 f \bar{f} = P_\mu P^\mu = P^2\, ,
\end{equation}
and one obtains
\begin{equation}\label{6noe}
W^\mu W_\mu = -P^2 {\mathbf{s}}^2 \, .
\end{equation}

We determine now the form of the four algebraic constraints $T_a$
($a=1,2,3,4$) present in the action (\ref{act}).

The variables $\lambda^a(\tau)$ are Lagrange multipliers for the
four physical constraints $T_a$.  We take the physical constraints
in the following form\footnote{The justification of the form of the
constraints $T_a$ can be obtained by considering the symmetries of
the action (\ref{act}). It appears that the choice
(\ref{mass})-(\ref{Q}) and the interpretation of $s_i$ (see
(\ref{lagva2.18})) as covariant spin projection is
 related with the formulae for the corresponding Noether charges.}
\bl
\begin{eqnarray}
T_1\,: & \qquad &T \equiv 4f\bar f-m^2\approx 0 \, , \label{mass}\\
T_2\,: & \qquad &S \equiv {\mathbf{s}}^2-s(s+1)\approx 0 \, , \label{S}\\
T_3\,: & \qquad &S_3 \equiv s_3-m_3\approx 0 \, , \label{S-3}\\
T_4\,: & \qquad &Q \equiv s_0-q\approx 0 \, . \label{Q}
\end{eqnarray}\el
The real quantities ${\mathbf{s}}=(s_r)= (s_1, s_2, s_3)$ and $s_0$
which are present in (\ref{S})-(\ref{Q}) are defined in terms of the
Lagrangian variables $s_k^j$ as follows
\begin{equation}\label{lagva2.18}
s_0={\frac{1}{2}}s_k{}^k\, ,\qquad s_r={\frac{1}{2}}s_k{}^j
(\tau_r)_j{}^k \, ,\qquad r=1,2,3\, ,
\end{equation}
where $(\tau_r)_j{}^k$ are the  Pauli matrices.

The constraint (\ref{mass}) defines the mass $m$ of the particle
because using it together with (\ref{5bisnoe}) we obtain that
\begin{equation}\label{PmuPmu}
P_\mu P^\mu= m^2\, .
\end{equation}
The constraints (\ref{S}) and (\ref{S-3}) are introduced in the
action (\ref{act}) in order to obtain a definite spin $s$ and the
covariant spin projection $s_3$ whereas the constraint (\ref{Q})
defines the $U(1)$ charge $q$ of the particle.

In the subsection \ref{Sect.TimeEvolution} we shall see, from the
preservation of the constraints in time, that secondary constraints
do not appear in our model. Thus, the full set of constraints is
given by the physical constraints (\ref{mass})-(\ref{Q}) and by the
primary ones (\ref{c:p-x})-(\ref{c:p-s}).

\subsection{Analysis of the primary constraints}

If we transform the  twelve constraints (\ref{c:p-x}),
(\ref{c:p-pi}) to equivalent Lorentz-invariant expressions by
contracting them with the spinors $\pi_{\alpha k}$ and $\bar
\pi_{\dot\alpha}^k$ (the matrices
$\pi_{\alpha}^{k},\opi_\dalpha^{\phantom{\dalpha} k}$ are invertible
due to (\ref{pipi1})-(\ref{pipi4})) the discussion of the
constraints is simplified and their splitting into first and second
class is clearer. After such contractions, the eight expressions
(\ref{c:p-pi}) take the form
\begin{equation}
\pi_{\alpha k}P^{\alpha j}- \frac{i}{2}s_k{}^j \approx 0\quad
,\qquad \bar P^{\dot\alpha}_j \bar \pi_{\dot\alpha}^k
+\frac{i}{2}s_j{}^k \approx 0\, .
\end{equation}
By considering the sum and the difference of the  expressions above,
we obtain the following set of eight constraints
\begin{equation}\label{D,B}
D_k{}^j \equiv {\cal D}_k{}^j +s_k{}^j \approx 0\, , \qquad
B_k{}^j \equiv {\cal B}_k{}^j \approx 0\, ,
\end{equation}
where the quantities
\begin{equation}\label{calD,B}
{\cal D}_k{}^j \equiv i(\pi_{\alpha k}P^{\alpha j} - \bar
P^{\dot\alpha}_k \bar \pi_{\dot\alpha}^j) \, , \qquad {\cal
B}_k{}^j \equiv i(\pi_{\alpha k}P^{\alpha j} + \bar P^{\dot\alpha}_k
\bar \pi_{\dot\alpha}^j)\, ,
\end{equation}
contain only spinorial phase space variables.

The  four constraints (\ref{c:p-x}), after
contraction with spinors, take the form
\begin{equation}\label{Px}
C_k{}^l \equiv {\cal P}_k{}^l +m^2\delta_k{}^l \approx 0\, ,
\end{equation}
where we  take into account (\ref{5bisnoe}) and (\ref{p-x}) and
introduce the following notation
\begin{equation}\label{calP}
{\cal P}_k{}^l \equiv 4 \pi_{\alpha
k}P^{\alpha\dot\beta}\bar\pi_{\dot\beta}^{\phantom{\beta}l}\, .
\end{equation}
Note that the spinorial bilinears introduce an
orthogonal basis ${\cal P}_{\mu \, k}^{\ \ \ l}$ defined by
(\ref{3noe}) or equivalently
\begin{equation}
P_{\mu}^{(r)}=(\tau^{(r)}){}_{l}{}^{k} {\cal P}_{\mu \, k}^{ \ \ \
l} = (\tau^{(r)}){}_{l}{}^{k}\, \pi_{\phantom{\alpha}k}^{\alpha}
 (\sigma_\mu)_{\alpha \dot{\beta}}
\, {\bar{\pi}}^{\dot{\beta}l}\, ,
\end{equation}
with $P_{\mu\ l}^{\phantom{\mu\ l}l}\equiv P_\mu$. Then, the
relation (\ref{Px}) can be written also as follows
\begin{equation}
\mathcal{P}_k^{\phantom{k}l}=2P_{\mu\ k}^{\phantom{\mu\
k}l}P^\mu=-m^2\delta_k^l\, .
\end{equation}

The algebra of the constraints (\ref{D,B}), (\ref{Px}) and
(\ref{c:p-s}) becomes more transparent if we introduce the following
$SU(2)$ scalar and vector quantities:\bl
\begin{equation}\label{D,B-r,0}
D_r ={\frac{1}{2}}D_k{}^j(\tau_r)_j{}^k\, ,\quad D_0
={\frac{1}{2}}D_k{}^k\quad ;\quad B_r ={\frac{1}{2}}B_k{}^j
(\tau_r)_j{}^k\, ,\quad B_0 ={\frac{1}{2}}B_k{}^k\quad ;
\end{equation}
\begin{equation}\label{calD,B-r,0}
{\cal D}_r ={\frac{1}{2}}{\cal D}_k{}^j(\tau_r)_j{}^k\, ,\quad
{\cal D}_0 ={\frac{1}{2}}{\cal D}_k{}^k\quad ;\quad {\cal B}_r
={\frac{1}{2}}{\cal B}_k{}^j (\tau_r)_j{}^k\, ,\quad {\cal B}_0
={\frac{1}{2}}{\cal B}_k{}^k\quad ;
\end{equation}
\begin{equation}\label{calP-r,0}
{\cal P}_r ={\frac{1}{2}}{\cal P}_k{}^j(\tau_r)_j{}^k\, ,\quad
{\cal P}_0 ={\frac{1}{2}}{\cal P}_k{}^k\quad ;
\end{equation}
\begin{equation}
C_r ={\frac{1}{2}}C_k{}^j(\tau_r)_j{}^k\, ,\quad C_0
={\frac{1}{2}}C_k{}^k\quad ;
\end{equation}
\begin{equation}\label{s,P-r,0}
{\cal P}_{(s)}{}_r ={\frac{1}{2}} P_{(s)}{}_k{}^j
(\tau_r)_j{}^k\, ,\quad P_{(s)}{}_0 ={\frac{1}{2}}{\cal
P}_{(s)}{}_k{}^k\, ,
\end{equation}\el
where $(\tau_r)_j{}^k$, $r=1,2,3$ are the isospin Pauli matrices. In
terms of the variables (\ref{D,B-r,0})-(\ref{s,P-r,0}) and
(\ref{lagva2.18}) the constraints (\ref{D,B}), (\ref{Px}) and
(\ref{c:p-s}) take the form\bl
\begin{eqnarray}
R_r \equiv P_{(s)}{}_r \approx 0\, , & \qquad &R_0 \equiv
P_{(s)}{}_0
\approx 0\, , \label{const-Ps}\\
D_r \equiv {\cal D}_r +s_r \approx 0\, , & \qquad &D_0 \equiv
{\cal
D}_0+s_0 \approx 0\, , \label{const-D}\\
B_r \equiv {\cal B}_r \approx 0\, , & \qquad &B_0 \equiv {\cal
B}_0
\approx 0\, , \label{const-B}\\
C_r \equiv {\cal P}_r \approx 0 \, , & \qquad &C_0 \equiv {\cal
P}_0 +m^2 \approx 0\, . \label{const-Px}
\end{eqnarray}\el
Thus, our full set of the constraints is described now by the four
physical constraints (\ref{mass})-(\ref{Q}) and by the sixteen
primary constraints (\ref{const-Ps})-(\ref{const-Px}).

We present now the canonical Poisson brackets of the coordinates
(\ref{coord}) and their momenta (\ref{p-x})-(\ref{p-s}) \bl
\begin{eqnarray}
\{ x^\mu, P_\nu\}=\delta^\mu_\nu \, ,&\qquad &\{ s_k{}^j,
P_{(s)}{}^n{}_l\}=\delta_k^n\delta^j_l
\, ,\label{CBxP}\\
\{ \pi_{\alpha k}, P^{\beta j}\}=\delta_\alpha^\beta \delta_k^j
\, ,&\qquad& \{
{\bar\pi}_{\dot\alpha}^k, \bar P^{\dot\beta}_j\}= \delta_{\dot\alpha}^{\dot\beta}\delta^k_j \, ,\\
\{ s_0, P_{(s)}{}_0\}=\frac{1}{2}\, ,&\qquad&\{ s_r,
P_{(s)}{}_q\}=\frac{1}{2}\delta_{rq}\, .\label{CBsrPs}
\end{eqnarray}
\el
These allow us to compute the Poisson brackets between the
quantities (\ref{calD,B-r,0})-(\ref{calP-r,0}). The the
non-vanishing ones are\bl
\begin{equation}\label{calDB}
\{ {\cal D}_r, {\cal D}_p\}= -\epsilon_{rpq} {\cal D}_q \, ,\quad
\{ {\cal D}_r, {\cal B}_p\}= -\epsilon_{rpq} {\cal B}_q \, ,\quad
\{ {\cal B}_r, {\cal B}_p\}= -\epsilon_{rpq} {\cal D}_q \, ,
\end{equation}
\begin{equation}\label{calDBP}
\{ {\cal P}_r, {\cal D}_p\}= -\epsilon_{rpq} {\cal P}_q \, ,\quad
\{ {\cal P}_r, {\cal B}_p\}= i\delta_{rp} {\cal P}_0 \, ,\quad \{
{\cal P}_0, {\cal B}_r\}= i{\cal P}_r \, ,
\end{equation}
\begin{equation}\label{calBPf}
\{ {\cal P}_r, {\cal B}_0\}=i {\cal P}_r \, ,\quad \{ {\cal P}_0,
{\cal B}_0\}= i {\cal P}_0 \, .
\end{equation}\el
{}From (\ref{calDB}) we see that the three quantities
$\mathcal{D}_r$ are the generators of $SO(3)$ and the three
quantities $\mathcal {B}_r$ extend the $\mathcal{SO}(3)$ algebra to
the Lorentz symmetry $SO(3,1)\simeq sl(2;\mathbb{C})$. Because the
generators $\mathcal{D}_r$, $\mathcal{B}_r$ are scalars we shall
call them internal symmetry generators.

{}From (\ref{calDBP}), (\ref{calBPf}) we see that the quantities
${\cal P}_0$, ${\cal P}_r$ which describe the covariant projections
of the four-momentum on the  composite four-vectors (\ref{3noe})
extend the internal Lorentz generators $({\cal D}_r, {\cal B}_r)$ to
an internal Poincar\'{e} algebra.

{}Finally, we can write the complete list of non-vanishing Poisson
brackets between all twenty constraints in our model  (four physical
constraints (\ref{mass})-(\ref{Q}) and sixteen primary ones
(\ref{const-Ps})-(\ref{const-Px})): \bl
\begin{eqnarray}
\{D_r, D_p\} & = & -\epsilon_{rpq}D_q +\epsilon_{rpq}s_q\, , \label{DD}\\
\{D_r, B_p\} & = & -\epsilon_{rpq}B_q \, , \label{DB}\\
\{B_r, B_p\} & = & -\epsilon_{rpq}D_q +\epsilon_{rpq}s_q\, ,
\label{BB}
\end{eqnarray}
\begin{eqnarray}
\{C_r, D_p\} & = & -\epsilon_{rpq}C_q \, , \label{CD}\\
\{C_r, B_p\} & = & i\delta_{rp}C_0 -i\delta_{rp}m^2\, , \label{CB}\\
\{C_r, B_0\} & = & iC_r \, , \label{CB0}
\end{eqnarray}
\begin{eqnarray}
\{C_0, B_r\} & = & iC_r \, , \label{C0B}\\
\{C_0, B_0\} & = & iC_0 -im^2\, , \label{C0B0}
\end{eqnarray}
\begin{equation}\label{TB}
\{T, B_0\} = 2iT +2im^2\, ,
\end{equation}
\begin{eqnarray}
\{D_r, R_p\} & = & {\frac{1}{2}}\delta_{rp} \, , \label{DR}\\
\{D_0, R_0\} & = & {\frac{1}{2}}\, , \label{D0R0}
\end{eqnarray}
\begin{eqnarray}
\{S, R_p\} & = & s_p \, , \label{SR}\\
\{S_3, R_3\} & = & {\frac{1}{2}}\, , \label{S3R}\\
\{Q, R_0\} & = & {\frac{1}{2}} \, . \label{QR}
\end{eqnarray}\el

\subsection{Time evolution of constraints and their split into first and second
class}\label{Sect.TimeEvolution}

The action (\ref{act}) is invariant under an arbitrary rescaling
on the world line $\tau\rightarrow \tau'=\tau'(\tau)$ and the
canonical Hamiltonian vanishes
\begin{equation}
\mathcal{H}=\mathcal{P}_L \dot{{Q}}_L -\mathcal{L}=0\, .
\end{equation}
The total Hamiltonian is given, therefore, by a linear combination
of all the constraints
\begin{eqnarray}
{\cal H}^C & = & \lambda^{(D)}_r D_r+ \lambda^{(D)}_0 D_0+
\lambda^{(B)}_r B_r+ \lambda^{(B)}_0 B_0+ \lambda^{(C)}_r C_r+
\lambda^{(C)}_0 C_0+ \nonumber\\
&& +\lambda^{(R)}_r R_r+ \lambda^{(R)}_0 R_0+ \lambda^{(T)} T+
\lambda^{(S)} S+ \lambda^{(S_3)} S_3+ \lambda^{(Q)} Q\, .
\label{H}
\end{eqnarray}

Imposing   the preservation of all the constraints in time (see
Appendix \ref{Appendix}) we find that four out of twenty Lagrange
multipliers are not determined. The Hamiltonian (\ref{H}) takes the
following final form
\begin{equation}\label{H1}
{\cal H} = \lambda^{(C)}_0 {\cal F}+ \lambda^{(S)} {\cal S}+
\lambda^{(S_3)} {\cal S}_3+ \lambda^{(Q)} {\cal Q}\, ,
\end{equation}
where\bl
\begin{equation}\label{cF}
{\cal F} = C_0 + {\frac{1}{2}}T \simeq 0 \, ,
\end{equation}
\begin{equation}\label{cS}
{\cal S} = S- 2s_r D_r \simeq 0 \, ,
\end{equation}
\begin{equation}\label{cS3}
{\cal S}_3 = S_3-D_3- 2\epsilon_{3rq}s_q R_r \simeq 0 \, ,
\end{equation}
\begin{equation}\label{cQ}
{\cal Q} = Q-D_0 \simeq 0
\end{equation}\el
describe the  four first class constraints. The other  sixteen
constraints can be presented as eight pairs of canonically
conjugated second class constraints\bl
\begin{eqnarray}
D_r \equiv {\cal D}_r +s_r \approx 0 & \quad\Leftrightarrow\quad &
R_r \equiv P_{(s)}{}_r \approx 0\, , \label{pDR}\\
D_0 \equiv {\cal D}_0+s_0 \approx 0 & \quad\Leftrightarrow\quad &
R_0 \equiv P_{(s)}{}_0 \approx 0\, , \label{pDR0}\\
B_r \equiv {\cal B}_r \approx 0 & \quad\Leftrightarrow\quad & C_r
\equiv {\cal P}_r \approx 0\, , \label{pBC}\\
B_0 \equiv {\cal B}_0 \approx 0 & \quad\Leftrightarrow\quad & T
\approx 0\, . \label{pBT}
\end{eqnarray}\el

The subset of constraints $D_r$, $B_r$ does not close under the PB
operation (see Eqs. (\ref{DD})-(\ref{BB})). This can be avoided if
we introduce the following linear combination of constraints\bl
\begin{equation}\label{D-pr}
D_r{}^\prime \equiv D_r-\epsilon_{rpq}s_pR_q={\cal D}_r +s_r
-\epsilon_{rpq}s_p{\cal P}_{(s)}{}_q \approx 0\, ,
\end{equation}
\begin{equation}\label{B-pr}
B_r{}^\prime \equiv B_r+ {\frac{i}{2m^2}}\epsilon_{rpq}s_p C_q={\cal
B}_r+ {\frac{i}{2m^2}}\epsilon_{rpq}s_p {\cal P}_q \approx 0\, .
\end{equation}\el
These have the following Poisson brackets\bl
\begin{equation}
\{D_r{}^\prime, D_p{}^\prime\} = -\epsilon_{rpq}D_q{}^\prime
+{\frac{1}{2}}(s_r R_p-s_p R_r)\, ,
\end{equation}
\begin{equation}
\{D_r{}^\prime, B_p{}^\prime\} = -\epsilon_{rpq}B_q{}^\prime
+{\frac{i}{4m^2}}(\delta_{rp}s_q C_q-s_p C_r)\, ,
\end{equation}
\begin{equation}
\{B_r{}^\prime, B_p{}^\prime\} = -\epsilon_{rpq}D_q{}^\prime -(s_r
R_p-s_p R_r) +{\frac{1}{m^2}}\epsilon_{rpq}s_q C_0 \, ,
\end{equation}\el
which vanish on the surface of the constraints.

\section{Solving the second class
constraints}\label{Sect.Sol.2nd.Class.Constr}
\setcounter{equation}{0}

{}From the relations (\ref{DR})-(\ref{D0R0}) we see that the four
pairs of constraints described by Eqs. (\ref{pDR})-(\ref{pDR0})
satisfy canonical PB, {\it i.e.} they have the so-called resolution
form\footnote{A pair of constraints $A\approx 0,B\approx 0$ have the
resolution form in the phase space $(x_i,p_i)$ $i=1,\dots,N$ if they
have the form given by the following formulae:
\begin{equation}
A=x_1-f(x_r,p_r)\approx 0\qquad B=p_1\approx 0\quad (r=2,3,\dots,N)
\end{equation}
This form of the constraints was considered by Dirac \cite{Dirac}.
In such a case the Dirac brackets are identical with the canonical
PB.}. Therefore, if we exclude the variables $s_r$, $s_0$ and
$P_{(s)}{}_r$, $P_{(s)}{}_0$ by means of the constraints
(\ref{pDR}), (\ref{pDR0}) the Dirac brackets for the remaining
variables will coincide with the canonical ones (see Eqs.
(\ref{CBxP})-(\ref{CBsrPs})). Consequently, in all expressions we
should insert
\begin{equation}\label{s}
s_r = - {\cal D}_r \, ,\qquad s_0=- D_0 \, ,\qquad P_{(s)}{}_r
= 0 \, ,\qquad P_{(s)}{}_0 =0 \, .
\end{equation}

In order to treat the three pairs of second class constraints
(\ref{pBC}) it is convenient to introduce in the phase space
$(Q_L,\mathcal{P}_L)$ the new variables canonically conjugated to
the constraints $\mathcal{P}_r$. In such a way the constraints
(\ref{pBC}) will also have the resolution form, and the introduction
of corresponding Dirac brackets   will not change the PB relations
for the other variables. For this purpose we pass from the
twenty-four initial canonical variables (we recall that $s_r,s_o$
and $P_{(s)r}, P_{(s)0}$ are not present due to (\ref{s}))
\begin{equation*}
(x^\mu\, ; P_\mu)\, , \quad (\pi_{\alpha k}\, ; P^{\alpha
k})\, , \quad (\bar\pi_{\dot\alpha}^k\, ; \bar
P^{\dot\alpha}_k)\, ,
\end{equation*}
to the new  twenty-four canonical variables
\begin{equation*}
(\widetilde{x}_0, \widetilde{x}_r\, ; {\cal P}_0 , {\cal P}_r)\quad
, \quad (\pi^\prime_{\alpha k}\, ; {\cal P}^{\alpha k})\, , \quad
(\bar\pi^{\prime k}_{\dot\alpha}\,; \bar{\cal P}^{\dot\alpha}_k)
\, ,
\end{equation*}
where the variables ${\cal P}_0$, ${\cal P}_r$ are given by the
expressions (\ref{calP}), (\ref{calP-r,0}) describing covariant
momentum projections, {\it i.e.}
\begin{equation}\label{calP1}
{\cal P}_0 = \pi_{\alpha
k}\sigma_\mu^{\alpha\dot\beta}\bar\pi_{\dot\beta}^k P^\mu\, ,
\qquad {\cal P}_r = (\tau_r)_j{}^k\pi_{\alpha
k}\sigma_\mu^{\alpha\dot\beta}\bar\pi_{\dot\beta}^j P^\mu\, .
\end{equation}
 Besides, we take
\begin{equation}\label{pi-prime}
\pi^\prime_{\alpha k} = \pi_{\alpha k}\, , \qquad \bar\pi^{\prime
k}_{\dot\alpha} = \bar\pi^k_{\dot\alpha}\, .
\end{equation}
and hence we will omit the prime in the transformed spinors. One can
check that the new coordinates and spinorial momenta, which satisfy
canonical commutation relations\bl
\begin{equation}\label{newPBbis}
\{ \widetilde{x}_0, {\cal P}_0\}=1\, ,\quad \{ \widetilde{x}_r,
{\cal P}_q\}=-\delta_{rq} \, ,\quad
\end{equation}
\begin{equation}\label{newPB}
\{ \pi^\prime_{\alpha k}, {\cal P}^{\beta j}\}=\delta_\alpha^\beta
\delta_k^j \, ,\quad \{ \bar\pi^{\prime k}_{\dot\alpha},
\bar{\cal P}^{\dot\beta}_j\}=
\delta_{\dot\alpha}^{\dot\beta}\delta^k_j\, ,
\end{equation}\el
are given by the formulae\footnote{The generating function of this
canonical transformation has the form
$$
F(P^\mu, \pi_{\alpha k}, \bar\pi_{\dot\beta}^k; \widetilde{x}_0,
\widetilde{x}_r, {\cal P}^{\alpha k}, \bar{\cal
P}^{\dot\alpha}_k)=
$$
$$
= -[\pi_{\alpha k}\sigma_\mu^{\alpha\dot\beta}\bar\pi_{\dot\beta}^k
P^\mu]\widetilde{x}_0 + [(\tau_r)_j{}^k\pi_{\alpha
k}\sigma_\mu^{\alpha\dot\beta}\bar\pi_{\dot\beta}^j P^\mu]
\widetilde{x}_r + \pi_{\alpha k}{\cal P}^{\alpha k}
+\bar\pi_{\dot\alpha}^k \bar{\cal P}^{\dot\alpha}_k \, .
$$
In this generating function there are encoded the expressions
(\ref{calP1}), (\ref{pi-prime}) by ${\cal P}_0 = -\frac{\partial
F}{\partial \widetilde{x}_0}$, ${\cal P}_r = \frac{\partial
F}{\partial \widetilde{x}_r}$, $\pi^\prime_{\alpha k} =
\frac{\partial F}{\partial {\cal P}^{\alpha k}}$, $\bar\pi^{\prime
k}_{\dot\alpha} = \frac{\partial F}{\partial \bar{\cal
P}^{\dot\alpha}_k}$. From $x_\mu = -\frac{\partial F}{\partial
P^\mu}$, $ P^{\alpha k} = \frac{\partial F}{\partial \pi_{\alpha
k}}$, $\bar P^{\dot\alpha}_k = \frac{\partial F}{\partial
\bar\pi_{\dot\alpha}^k}$ we obtain the expressions (\ref{x}),
(\ref{calP-new}). }
\begin{equation}\label{x}
\widetilde{x}_0={\frac{1}{2|f|^2}} \pi^{\alpha k}x_{\alpha\dot\beta}
\bar\pi^{\dot\beta}_k \, , \qquad
\widetilde{x}_r=-{\frac{1}{2|f|^2}}(\tau_r)_k{}^j \pi^{\alpha
k}x_{\alpha\dot\beta} \bar\pi^{\dot\beta}_j\, ,
\end{equation}
\begin{equation}\label{calP-new}
{\cal P}^{\alpha k} =P^{\alpha k} - {\frac{2}{\bar f}} \pi^{\gamma\,
k} x_{\gamma\dot\beta} P^{\dot\beta\alpha}\, , \qquad \bar{\cal
P}^{\dot\alpha}_k =\bar P^{\dot\alpha}_k + {\frac{2}{f}}
P^{\dot\alpha\beta} x_{\beta\dot\gamma} \bar\pi^{\dot\gamma}_k \quad
.
\end{equation}
We see from (\ref{x}) that the new covariant spacetime coordinates
$(\widetilde{x}_0,\widetilde{x}_r)$ are described by four covariant
projections on the composite four-vectors (\ref{3noe}) Using Eq.
(\ref{calP-new}) we obtain the following useful relations \bl
\begin{eqnarray}
\pi_{\alpha k}P^{\alpha k} &=& \pi_{\alpha k}{\cal P}^{\alpha k} -
\widetilde{x}_0{\cal P}_0 + \widetilde{x}_r{\cal P}_r \, ,\\
\bar P^{\dot\alpha}_k \bar\pi_{\dot\alpha}^k&=& \bar{\cal
P}^{\dot\alpha}_k \bar\pi_{\dot\alpha}^k - \widetilde{x}_0{\cal
P}_0 +
\widetilde{x}_r{\cal P}_r \, ,\\
(\tau_r)_l{}^k\pi_{\alpha k}P^{\alpha l} &=& (\tau_r)_l{}^k
\pi_{\alpha k}{\cal P}^{\alpha l}
- \widetilde{x}_0{\cal P}_r + x_r{\cal P}_0 -i\epsilon_{rpq}\widetilde{x}_p{\cal P}_q\, ,\\
(\tau_r)_l{}^k\bar P^{\dot\alpha}_k \bar\pi_{\dot\alpha}^l&=&
(\tau_r)_l{}^k\bar{\cal P}^{\dot\alpha}_k \bar\pi_{\dot\alpha}^l-
\widetilde{x}_0{\cal P}_r + \widetilde{x}_r{\cal P}_0
+i\epsilon_{rpq}\widetilde{x}_p{\cal P}_q\, .
\end{eqnarray}
\el In terms of the new variables the expressions (\ref{calD,B})
contain an additional term depending on $\widetilde{x}_0$,
$\widetilde{x}_r$, ${\cal P}_0$, ${\cal P}_r$, namely \bl
\begin{equation}\label{B-}
{\cal B}_0 \rightarrow {\cal B}_0 - i\widetilde{x}_0{\cal P}_0 +
i\widetilde{x}_r{\cal P}_r\, , \qquad {\cal B}_r \rightarrow
{\cal B}_r - i\widetilde{x}_0{\cal P}_r + i\widetilde{x}_r{\cal
P}_0\, ,
\end{equation}
\begin{equation}\label{D-}
{\cal D}_0 \rightarrow {\cal D}_0 \, , \qquad {\cal D}_r
\rightarrow {\cal D}_r + \epsilon_{rpq}\widetilde{x}_p{\cal
P}_q\, ,
\end{equation}\el
where in the {\it r.h.s.} of these relations the terms ${\cal B}_0$,
${\cal B}_r$, ${\cal D}_0$, ${\cal D}_r$ are obtained from
(\ref{calD,B}) by the replacements $P^{\alpha k}\rightarrow{\cal
P}^{\alpha k}$, $\bar P^{\dot\alpha}_k\rightarrow\bar {\cal
P}^{\dot\alpha}_k$.

Thus, in the new variables $(\widetilde{x}_0, \widetilde{x}_r\, ,
{\cal P}_0 , {\cal P}_r\, ,\, \pi^\prime_{\alpha k}\, , {\cal
P}^{\alpha k}\, , \, \bar\pi^{\prime k}_{\dot\alpha}\,, \bar{\cal
P}^{\dot\alpha}_k)$ the three $B_r$ constraints in (\ref{pBC}) take
the form
\begin{equation}\label{Br}
B_r = {\cal B}_r- i\widetilde{x}_0{\cal P}_r + i\widetilde{x}_r{\cal
P}_0 \approx 0\, .
\end{equation}
The pairs of constraints (\ref{pBC}) satisfy the canonical PB,
rescaled by $\mathcal{P}_0$ (see (\ref{calDBP})), what is a trivial
extension of the resolution form. If we introduce Dirac brackets
consistent with the constraints (\ref{pBC}) one can exclude the
variables $\widetilde{x}_r$ and $\mathcal{P}_r$ by setting
\begin{equation}
\widetilde{x}_r =-\frac{i}{{\cal P}_0}{\cal B}_r\, , \qquad {\cal
P}_r =0\, ,
\end{equation}
again without any modification of the PB for the remaining variables
\sloppy $(\widetilde{x}_0,\mathcal{P}_0, \pi_{\alpha i}, \mathcal{P}_{\alpha
i},
\opi_\dalpha^{\phantom{\alpha}i},\overline{\mathcal{P}}^\dalpha_{\phantom{\alpha}i})$.

The only two remaining  second class constraints  have the
form\bl
\begin{eqnarray}
T=4f\of -m^2&=&0\, ,\label{T.Constraint}\\
B_0=\mathcal{B}_0-i\widetilde{x}_0\mathcal{P}_0&=&0\quad
.\label{B0.Constraint}
\end{eqnarray}\el
Subsequently, we introduce the Dirac brackets (DB) as follows
\begin{equation}\label{DB2}
\{y, y'\}_{D}=\{y,y'\}+\{y, B_0\}{\frac{i}{2(T+m^2)}}\{T, y'\} -\{y,
T\}{\frac{i}{2(T+m^2)}}\{B_0, y'\}\, ,
\end{equation}
where $y,y'\in Y_R=(\widetilde{x}_0,\mathcal{P}_0,\pi_{\alpha k},
\mathcal{P}^{\alpha
k},\opi_\dalpha^{\phantom{\alpha}k},
\overline{\mathcal{P}}^\dalpha_{\phantom{\alpha}k})$; $R=1,2, \ldots 18$.
It is easy to check that $\{y,T\}\neq 0$ only if
$y\in(\mathcal{P}^{\alpha
k},\overline{\mathcal{P}}^\dalpha_{\phantom{\alpha}k})$, {\it i.e.},
only those canonical PB that include the spinorial momenta are
modified. We obtain from (\ref{DB2}) the following explicit Dirac
brackets for the phase variables in $Y_R$  (we
present only the non-vanishing ones):
\bl
\begin{equation}\label{x-P}
\{x_0, {\cal P}_0\}_{D}=1 \, ,
\end{equation}
\begin{equation}\label{x-Pi}
\{x_0 , {\cal P}^{\beta j}\}_{D}={\frac{2f}{m^2}}x_0 \pi^{\beta
j}\, , \qquad \{x_0 , \bar {\cal
P}^{\dot\beta}_j\}_{D}=-{\frac{2\bar f}{m^2}}x_0
\bar\pi_{j}^{\dot\beta} \, ,
\end{equation}
\begin{equation}\label{P-Pi}
\{{\cal P}_0 , {\cal P}^{\beta j}\}_{D}=-{\frac{2f}{m^2}}{\cal P}_0
\pi^{\beta j}\, , \qquad \{{\cal P}_0 , \bar {\cal
P}^{\dot\beta}_j\}_{D}={\frac{2\bar f}{m^2}}{\cal P}_0
\bar\pi_{j}^{\dot\beta} \, ,
\end{equation}
\begin{equation}\label{pi-Pi}
\{\pi_{\alpha k} , {\cal P}^{\beta
j}\}_{D}=\delta_{\alpha}^{\beta}\delta_{k}^{j}
-{\frac{f}{m^2}}\pi_{\alpha k} \pi^{\beta j}\, , \qquad \{
\pi_{\alpha k} , \bar {\cal P}^{\dot\beta}_j\}_{D}={\frac{\bar
f}{m^2}}\pi_{\alpha k} \bar\pi_{j}^{\dot\beta} \, ,
\end{equation}
\begin{equation}\label{bpi-Pi}
\{\bar\pi^{k}_{\dot\alpha} , \bar {\cal P}^{\dot\beta}_j\}_{D}=
\delta_{\dot\alpha}^{\dot\beta}\delta_{j}^{k}+ {\frac{\bar
f}{m^2}}\bar\pi^{k}_{\dot\alpha} \bar\pi_{j}^{\dot\beta} \quad
,\qquad \{\bar\pi^{k}_{\dot\alpha} , {\cal P}^{\beta
j}\}_{D}=-{\frac{f}{m^2}}\bar\pi^{k}_{\dot\alpha} \pi^{\beta j}\quad
,
\end{equation}
\begin{equation}\label{Pi-Pi}
\{{\cal P}^{\alpha k} , {\cal P}^{\beta j}\}_{D}=
-{\frac{f}{m^2}}(\pi^{\alpha k} {\cal P}^{\beta j}- \pi^{\beta j}
{\cal P}^{\alpha k})\, ,
\end{equation}
\begin{equation}\label{bPi-bPi}
\{ \bar {\cal P}^{\dot\alpha}_k , \bar {\cal
P}^{\dot\beta}_j\}_{D}={\frac{\bar f}{m^2}}(
\bar\pi_{k}^{\dot\alpha}\bar {\cal P}^{\dot\beta}_j
-\bar\pi_{j}^{\dot\beta}\bar {\cal P}^{\dot\alpha}_k )\, ,
\end{equation}
\begin{equation}\label{bPi-Pi}
\{{\cal P}^{\alpha k} , \bar {\cal P}^{\dot\beta}_j\}_{D}=
-{\frac{f}{m^2}}\pi^{\alpha k}\bar{\cal P}^{\dot\beta}_j -
{\frac{\bar f}{m^2}}\bar\pi_{j}^{\dot\beta}{\cal P}^{\alpha k}\quad
.
\end{equation}
\el The brackets (\ref{x-P})-(\ref{bPi-Pi}) are consistent with the
second class constraints (\ref{T.Constraint})-(\ref{B0.Constraint})
{\it i.e.} for all the variables $Y_R$ ($R=1 \ldots 18$) we have
\begin{equation}\label{bPi-Pi-bis}
\{ T, Y_R \}_D = \{ D, Y_R \}_D = 0
\end{equation}

We observe that the relation (\ref{T.Constraint}) reduces one
spinorial degree of freedom, {\it i.e.} we are left with seven
unconstrained spinorial coordinates.

\section{First quantization and solution
of the first class constraints}\label{Sect.Sol.1st.Class.Constr}
\setcounter{equation}{0}

\subsection{First class constraints}

After taking into account all the  sixteen second class constraints
(\ref{pDR})-(\ref{pBT}) there remain the following eighteen phase
space variables
\begin{equation}\label{newvariables}
\widetilde{x}_0\, ,\,\, {\cal P}_0\quad ;\qquad \pi_{\alpha
k}\, ,\,\, {\cal P}^{\alpha k}; \qquad
\bar\pi^{k}_{\dot\alpha}\, ,\,\, \bar{\cal P}^{\dot\alpha}_k\quad
,
\end{equation}
which are constrained by two algebraic relation
(\ref{T.Constraint}), (\ref{B0.Constraint}) and satisfy the Dirac
brackets (\ref{x-P})-(\ref{bPi-Pi}) (the remaining ones are
canonical). After performing the quantization of the canonical Dirac
brackets $\{y,y'\}_D\rightarrow\frac{1}{i}[\hat y,\hat y']$ (we put
$\hbar=1$) one obtains the corresponding commutation relations,
where we should keep the order of the quantized momenta as it is
written in the formulae (\ref{Pi-Pi})-(\ref{bPi-Pi}), {\it i.e.} we
use the `$qp$-ordering'.

The  sixteen independent degrees of freedom described by the
variables (\ref{newvariables}) are additionally restricted by the
four first class constraints (\ref{cF})-(\ref{cQ}). These, after the
use of some identities following from the second class constraints,
can be written in the following form\bl
\begin{equation}\label{1}
{\cal P}_0 +m^2 \approx 0\, ,
\end{equation}
\begin{equation}\label{2}
{\cal D}_r {\cal D}_r -s(s+1) \approx 0\, ,
\end{equation}
\begin{equation}\label{3}
{\cal D}_3 +m_3 \approx 0\, ,
\end{equation}
\begin{equation}\label{4}
{\cal D}_0 +q \approx 0\, ,
\end{equation}\el
where the numerical values of $m,s,m_3$ and $q$ describe mass, spin,
spin projection and internal Abelian (electric) charge.

\subsection{Covariant solution of the constraints}\label{Sect.Covar.Sol}

In order to write the first class constraints (\ref{1})-(\ref{4}) as
the wave equations we take the Schr\"{o}dinger realization of the
quantized variables (\ref{newvariables}) on the commuting
generalized coordinate space $(\widetilde{x}_0,\pi_{\alpha j},
\opi_{\dot\alpha}^j)$. The corresponding generalized momenta
$(\mathcal{P}_0,\mathcal{P}^{\beta
j},\overline{\mathcal{P}}^\dbeta_{\phantom{\beta}j})$ have the
following differential realizations:\bl
\begin{equation}\label{calP-0}
{\cal P}_0=-i\frac{\partial}{\partial \widetilde{x}_0}\, ,
\end{equation}
\begin{equation}\label{calP-a}
{\cal P}^{\beta j}=-i\frac{\partial}{\partial\pi_{\beta j}}
+i\frac{f}{m^2}\pi^{\beta j}\left(\pi_{\alpha
k}\frac{\partial}{\partial\pi_{\alpha k}} + \bar \pi_{\dot\alpha}^k
\frac{\partial}{\partial\bar \pi_{\dot\alpha}^k}
-2\widetilde{x}_0\frac{\partial}{\partial
\widetilde{x}_0}\right)\, ,
\end{equation}\
\begin{equation}\label{calP-b}
\bar {\cal
P}^{\dot\beta}_j=-i\frac{\partial}{\partial\bar\pi^{j}_{\dot\beta}}
-i\frac{\bar f}{m^2}\bar\pi_{j}^{\dot\beta}\left(\pi_{\alpha
k}\frac{\partial}{\partial\pi_{\alpha k}} + \bar \pi_{\dot\alpha}^k
\frac{\partial}{\partial\bar \pi_{\dot\alpha}^k}
-2\widetilde{x}_0\frac{\partial}{\partial
\widetilde{x}_0}\right)\, .
\end{equation}\el
The consistency of our quantization procedure can be obtained a
posteriori by  checking that the realizations
(\ref{calP-0})-(\ref{calP-b}) satisfy the commutation relations
obtained by the quantization of Dirac brackets
(\ref{x-P})-(\ref{bPi-Pi}). One can also show that the relations
(\ref{T.Constraint})-(\ref{B0.Constraint}) are satisfied.
Subsequently, we obtain the following simple differential
realizations of the operators $\mathcal{D}_r,\mathcal{D}_0$ defining
the three first class constraints (\ref{2})-(\ref{4}):\bl
\begin{equation}\label{calD-0}
{\cal D}_0 = \frac{1}{2}\left(\pi_{\alpha
k}\frac{\partial}{\partial\pi_{\alpha k}} - \bar \pi_{\dot\alpha}^k
\frac{\partial}{\partial\bar \pi_{\dot\alpha}^k}\right)\, ,
\end{equation}
\begin{equation}\label{calD-r}
{\cal D}_r = \frac{1}{2}(\tau_r)_j{}^k\left(\pi_{\alpha
k}\frac{\partial}{\partial\pi_{\alpha j}} - \bar \pi_{\dot\alpha}^j
\frac{\partial}{\partial\bar \pi_{\dot\alpha}^k}\right)\, ,
\qquad r=1,2,3
\end{equation}\el
The wavefunction has the following coordinate dependence
\begin{equation}\label{wf}
\Psi=\Psi(\widetilde{x}_0, \pi_{\alpha k}, \bar
\pi_{\dot\alpha}^k)\, .
\end{equation}
Substituting (\ref{calP-0}), (\ref{calD-0}) and (\ref{calD-r}) in
(\ref{1})-(\ref{4}) we obtain four generalized wave equations. We
shall solve them consecutively:

\medskip
\textit{i) Mass shell constraint (\ref{1}).}

The general solution of the constraint (\ref{1})
\begin{equation}
i\frac{\partial}{\partial \widetilde{x}_0}\Psi(\widetilde{x}_0,
\pi_{\alpha k}, \bar \pi_{\dot\alpha}^k)= m^2\Psi(\widetilde{x}_0,
\pi_{\alpha k}, \bar \pi_{\dot\alpha}^k)\, ,
\end{equation}
is the following
\begin{equation}\label{Psi}
\Psi(\widetilde{x}_0, \pi_{\alpha k}, \bar \pi_{\dot\alpha}^k)=
e^{-im^2 \widetilde{x}_0}\, \Phi(\pi_{\alpha k}, \bar
\pi_{\dot\alpha}^k)\, .
\end{equation}
Using the expression (\ref{x}) for $\widetilde{x}_0$ and
(\ref{c:p-x}) for $P_\mu$ and the constraint (\ref{mass}) we obtain
the following formula for the covariantized time coordinate
$\widetilde{\tau}=m\widetilde{x}_0$
\begin{equation}
m^2 \widetilde{x}_0 \equiv m\widetilde{\tau} =
 {\frac{m^2}{2|f|^2}} \pi^{\alpha k}
x_{\alpha\dot\beta} \bar\pi^{\dot\beta}_k = 2\pi^{\alpha k}
x_{\alpha\dot\beta} \bar\pi^{\dot\beta}_k =P^\mu x_\mu \, .
\end{equation}
Therefore, the exponent in the wavefunction (\ref{Psi}) has the
standard form of a plane wave
\begin{equation}\label{Psi-1}
\Psi(\widetilde{x}_0, \pi_{\alpha k}, \bar \pi_{\dot\alpha}^k)=
e^{-ix_\mu P^\mu}\, \Phi(\pi_{\alpha k}, \bar
\pi_{\dot\alpha}^k)\, ,
\end{equation}
where the four-momentum  $P_\mu$ is composite, {\it
i.e.}
\begin{equation}\label{Pmu}
P^\mu= \pi^{\alpha k}\sigma^\mu_{\alpha\dot\beta}
\bar\pi^{\dot\beta}_k\, .
\end{equation}

\textit{ii) Normalized spinors and  electric charge.}

The eight (real) variables $\pi_{\alpha k}$, $\bar
\pi_{\dot\alpha}^k$ define two variables $f$, $\bar f$ as follows
\begin{equation}\label{ff}
\pi^{\alpha k}\pi_{\alpha k}=2\bar f \, , \qquad
\bar\pi_{\dot\alpha k}\bar\pi^{\dot\alpha k} =2f
\end{equation}
and the remaining  six degrees of freedom  can be
described by the normalized spinors
\begin{equation}\label{u}
u_{\alpha i}= \left({{\frac{\bar f}{f}}}\right)^{-1/4}\pi_{\alpha
i}\, , \qquad \bar u_{\dot\alpha}^i= \overline{(u_{\alpha
i})}=\left({{\frac{\bar f}{f}}}\right)^{1/4}
\bar\pi_{\dot\alpha}^i\, .
\end{equation}
Due to the constraint (\ref{T.Constraint}) the modulus of $f$ is
given by the mass parameter $ |f|={{\frac{m}{2}}}$ and the variable
$y\in S^1$
\begin{equation}\label{y}
y\equiv {{\frac{\bar f}{f}}}\, ,
\end{equation}
defines the phase of $f$ which will be eliminated by the
constraint (\ref{4}).

{}From the definition (\ref{u}) the variables $u_{\alpha i}$, $\bar
u_{\dot\alpha}^i$ satisfy the relations
\begin{equation}\label{uu}
u^{\alpha k}u_{\alpha k}=2m \, , \qquad \bar u_{\dot\alpha k}\bar
u ^{\dot\alpha k} =2m\, .
\end{equation}
Three out of six degrees of freedom can be expressed via the formula
(\ref{Padb-2twistor}) by the components of the four-momentum vector.
One can also add that the variables $m^{-1/2}u_{\alpha i}$ form the
$SL(2,C)$ matrix of  and play the r\^{o}le of spinorial Lorentz
harmonics (see e.g. \cite{Ban90,FeZi95}).

Let us observe that using the expression (\ref{P(pi)}) for
$P_{\alpha\dot\alpha}=\pi_{\alpha}^{k}\bar\pi_{\dot\alpha k}$ we
obtain that the spinors $\pi_{\alpha k}$, $\bar \pi_{\dot\alpha}^k$
satisfy Dirac-type equations with complex mass $2f$
\begin{equation}\label{D-pi}
P_{\alpha\dot\alpha}\bar\pi^{\dot\alpha i}= f\pi_{\alpha}^i\, ,
\qquad P^{\dot\alpha\alpha}\pi_{\alpha i}= \bar
f\bar\pi^{\dot\alpha}_i\, .
\end{equation}
Substituting (\ref{u}) we obtain from (\ref{D-pi}) the standard
Dirac equations with real mass $m$ in two-component (Weyl) form
\begin{equation}\label{D-u}
P_{\alpha\dot\alpha}\bar u^{\dot\alpha i}= {{\frac{m}{2}}}
u_{\alpha}^i\, , \qquad P^{\dot\alpha\alpha} u_{\alpha i}=
{{\frac{m}{2}}}\bar u^{\dot\alpha}_i\, ,
\end{equation}
where $m$ is real because of (\ref{uu}).

Using the variables $y$, $u_{\alpha i}$, $\bar u_{\dot\alpha}^i$
(we recall that $|f|=\frac{m}{2}$) the differential operators
(\ref{calD-0}), (\ref{calD-r}) take the form\bl
\begin{equation}\label{calD-0-a}
{\cal D}_0 = 2y\frac{\partial}{\partial y} \, ,
\end{equation}
\begin{equation}\label{calD-r-a}
{\cal D}_r = \frac{1}{2}(\tau_r)_j{}^k\left(u_{\alpha
k}\frac{\partial}{\partial u_{\alpha j}} - \bar u_{\dot\alpha}^j
\frac{\partial}{\partial\bar u_{\dot\alpha}^k}\right)\, , \qquad
r=1,2,3
\end{equation}\el

The first class constraint (\ref{4}) looks as follows
\begin{equation}\label{calD-0-a1}
(2y\frac{\partial}{\partial y} +q) \Phi_m(y, u_{\alpha k}, \bar
u_{\dot\alpha}^k)=0\, ,
\end{equation}
and has the following solution
\begin{equation}\label{Phi-b}
\Phi_m(y, u_{\alpha k}, \bar u_{\dot\alpha}^k)= y^{-q/2}
\tilde\Phi_m(u_{\alpha k}, \bar u_{\dot\alpha}^k)\, ,
\end{equation}
where the function $\tilde\Phi_m(u_{\alpha k}, \bar
u_{\dot\alpha}^k)$ depends on the normalized spinors $u_{\alpha i}$
and $\bar u_{\dot\alpha}^i$ only.

\medskip
{\it iii) Spin description.}

Let us find  now the solution of the constraints
(\ref{2}), (\ref{3}) for the function $\tilde\Phi(u_{\alpha k}, \bar
u_{\dot\alpha}^k)$ using the polynomial expansion in spinor
variables
\begin{equation}\label{Phi}
\tilde\Phi(u_{\alpha k}, \bar u_{\dot\alpha}^k)=
\sum_{k,n=0}^{\infty} {\frac{1}{k!\,n!}}\, u_{\alpha_1 i_1}\ldots
u_{\alpha_k i_k} \bar u_{\dot\beta_1}^{j_1}\ldots\bar
u_{\dot\beta_n}^{j_n}\, \phi^{\alpha_1\ldots\alpha_k
\dot\beta_1\ldots\dot\beta_n}{}^{i_1\ldots i_k}_{j_1\ldots
j_n}(P_\mu)\, .
\end{equation}
The coefficient fields $\phi^{\alpha_1\ldots\alpha_k
\dot\beta_1\ldots\dot\beta_n}{}^{i_1\ldots i_k}_{j_1\ldots
j_n}(P_\mu)$ depend on $P_\mu= \sigma_{\mu\alpha\dot\beta}u^{\alpha
k}\bar u^{\dot\beta}_k= \sigma_{\mu\alpha\dot\beta}\pi^{\alpha
k}\bar\pi^{\dot\beta}_k$ and one can show that ${\cal D}_r
P_\mu={\cal D}_0 P_\mu=0$. The  wavefunctions $\phi$ in (\ref{Phi})
depending on the four-momentum are symmetric in all indices of the
same type
\begin{equation}
\phi^{\alpha_1\ldots\alpha_k
\dot\beta_1\ldots\dot\beta_n}{}^{i_1\ldots i_k}_{j_1\ldots
j_n}(P_\mu)= \phi^{(\alpha_1\ldots\alpha_k)
(\dot\beta_1\ldots\dot\beta_n)}{}^{(i_1\ldots i_k)}_{(j_1\ldots
j_n)}(P_\mu)\, ,
\end{equation}
and they do satisfy the following traceless condition
\begin{equation}
\phi^{\alpha_1\ldots\alpha_k \dot\beta_1\ldots\dot\beta_n}{}^{i
i_2\ldots i_k}_{i j_2\ldots j_n}(P_\mu)=0\, .
\end{equation}

Because $(\tau_r)_l{}^k(\tau_r)_i{}^j=2\delta_l^j \delta_i^k -
\delta_l^k \delta_i^j$ we obtain
\begin{equation*}\label{DrDr}
{\cal D}_r {\cal D}_r=\frac{1}{2}u_{\alpha
k}\frac{\partial}{\partial u_{\alpha l}} u_{\beta
l}\frac{\partial}{\partial u_{\beta k}} + \frac{1}{2}\bar
u_{\dot\alpha}^l \frac{\partial}{\partial\bar u_{\dot\alpha}^k}
\bar u_{\dot\beta}^k \frac{\partial}{\partial\bar u_{\dot\beta}^l}
- \bar u_{\dot\alpha}^l \frac{\partial}{\partial\bar
u_{\dot\alpha}^k} u_{\beta l}\frac{\partial}{\partial u_{\beta
k}}-
\end{equation*}
\begin{equation}
- \frac{1}{4} \left(u_{\alpha k}\frac{\partial}{\partial u_{\alpha
k}} - \bar u_{\dot\alpha}^k \frac{\partial}{\partial\bar
u_{\dot\alpha}^k}\right)^2\, .
\end{equation}
Using the following identities\bl
$$
\frac{1}{2} u_{\gamma k}\frac{\partial}{\partial u_{\gamma
l}}u_{\delta l}\frac{\partial}{\partial u_{\delta k}} u_{\alpha_1
i_1}\ldots u_{\alpha_k i_k} \bar u_{\dot\beta_1}^{j_1}\ldots\bar
u_{\dot\beta_n}^{j_n}\, \phi^{\alpha_1\ldots\alpha_k
\dot\beta_1\ldots\dot\beta_n}{}^{i_1\ldots i_k}_{j_1\ldots
j_n}(P_\mu)=
$$
\begin{equation}\label{DrDr1}
=\frac{k(k+1)}{2} u_{\alpha_1 i_1}\ldots u_{\alpha_k i_k} \bar
u_{\dot\beta_1}^{j_1}\ldots\bar u_{\dot\beta_n}^{j_n}\,
\phi^{\alpha_1\ldots\alpha_k
\dot\beta_1\ldots\dot\beta_n}{}^{i_1\ldots i_k}_{j_1\ldots
j_n}(P_\mu) \, ,
\end{equation}
$$
\frac{1}{2}\bar u_{\dot\gamma}^l \frac{\partial}{\partial\bar
u_{\dot\gamma}^k} \bar u_{\dot\delta}^k \frac{\partial}{\partial\bar
u_{\dot\delta}^l} u_{\alpha_1 i_1}\ldots u_{\alpha_k i_k} \bar
u_{\dot\beta_1}^{j_1}\ldots\bar u_{\dot\beta_n}^{j_n}\,
\phi^{\alpha_1\ldots\alpha_k
\dot\beta_1\ldots\dot\beta_n}{}^{i_1\ldots i_k}_{j_1\ldots
j_n}(P_\mu)=
$$
\begin{equation}\label{DrDr2}
=\frac{n(n+1)}{2} u_{\alpha_1 i_1}\ldots u_{\alpha_k i_k} \bar
u_{\dot\beta_1}^{j_1}\ldots\bar u_{\dot\beta_n}^{j_n}\,
\phi^{\alpha_1\ldots\alpha_k
\dot\beta_1\ldots\dot\beta_n}{}^{i_1\ldots i_k}_{j_1\ldots
j_n}(P_\mu) \, ,
\end{equation}
\begin{equation*}
\bar u_{\dot\gamma}^l \frac{\partial}{\partial\bar u_{\dot\gamma}^k}
u_{\delta l}\frac{\partial}{\partial u_{\delta k}} u_{\alpha_1
i_1}\ldots u_{\alpha_k i_k} \bar u_{\dot\beta_1}^{j_1}\ldots\bar
u_{\dot\beta_n}^{j_n}\, \phi^{\alpha_1\ldots\alpha_k
\dot\beta_1\ldots\dot\beta_n}{}^{i_1\ldots i_k}_{j_1\ldots
j_n}(P_\mu) \sim
\end{equation*}
\begin{equation}\label{DrDr3}
\sim \phi^{\alpha_1\ldots\alpha_k
\dot\beta_1\ldots\dot\beta_n}{}^{ii_2\ldots i_k}_{ij_2\ldots j_n}
(P_\mu)=0 \, ,
\end{equation}
$$
-\frac{1}{4} \left(u_{\gamma k}\frac{\partial}{\partial u_{\gamma
k}} - \bar u_{\dot\gamma}^k \frac{\partial}{\partial\bar
u_{\dot\gamma}^k}\right)^2 u_{\alpha_1 i_1}\ldots u_{\alpha_k i_k}
\bar u_{\dot\beta_1}^{j_1}\ldots\bar u_{\dot\beta_n}^{j_n}\,
\phi^{\alpha_1\ldots\alpha_k
\dot\beta_1\ldots\dot\beta_n}{}^{i_1\ldots i_k}_{j_1\ldots
j_n}(P_\mu)=
$$
\begin{equation}\label{DrDr4}
=-\frac{(k-n)^2}{4} u_{\alpha_1 i_1}\ldots u_{\alpha_k i_k} \bar
u_{\dot\beta_1}^{j_1}\ldots\bar u_{\dot\beta_n}^{j_n}\,
\phi^{\alpha_1\ldots\alpha_k
\dot\beta_1\ldots\dot\beta_n}{}^{i_1\ldots i_k}_{j_1\ldots
j_n}(P_\mu) \, ,
\end{equation}\el
we obtain the action of the operator ${\cal D}_r {\cal D}_r$ on the
polynomials in the expansion (\ref{Phi})
\begin{eqnarray}
{\cal D}_r {\cal D}_r u_{\alpha_1 i_1}\ldots u_{\alpha_k i_k} \bar
u_{\dot\beta_1}^{j_1}\ldots\bar u_{\dot\beta_n}^{j_n}\,
\phi^{\alpha_1\ldots\alpha_k
\dot\beta_1\ldots\dot\beta_n}{}^{i_1\ldots i_k}_{j_1\ldots j_n}(P_\mu)=\nonumber \\
={\frac{k+n}{2}}({\frac{k+n}{2}}+1) u_{\alpha_1 i_1}\ldots
u_{\alpha_k i_k} \bar u_{\dot\beta_1}^{j_1}\ldots\bar
u_{\dot\beta_n}^{j_n}\, \phi^{\alpha_1\ldots\alpha_k
\dot\beta_1\ldots\dot\beta_n}{}^{i_1\ldots i_k}_{j_1\ldots
j_n}(P_\mu)\, .\label{mon}
\end{eqnarray}

Thus, the solution of Eq. (\ref{2}) is
\begin{equation}\label{sol}
\tilde\Phi(u_{\alpha k}, \bar u_{\dot\alpha}^k)= \sum_{{k,n;\,
k+n=2s}} {\frac{1}{k!\,n!}}\, u_{\alpha_1 i_1}\ldots u_{\alpha_k
i_k} \bar u_{\dot\beta_1}^{j_1}\ldots\bar u_{\dot\beta_n}^{j_n}\,
\phi^{\alpha_1\ldots\alpha_k
\dot\beta_1\ldots\dot\beta_n}{}^{i_1\ldots i_k}_{j_1\ldots
j_n}(P_\mu)\, ,
\end{equation}
where in this expansion  only  spinorial polynomials of order $2s$
($k=k_1+k_2$, $n=n_1+n_2$) are present,
\begin{equation}\label{kns}
k+n=2s\, .
\end{equation}
where $k_i$ $(i=1,2)$ denotes the number of spinors $u^\alpha_i$,
and $n_i$ $(i=1,2)$ the number of spinors $\overline{u}_i^\dalpha$.

We can also derive the consequences of the relations (\ref{D-u})
between the spinors $u_{\alpha k}$, $\overline{u}_\alpha^k$.
Inserting these relations in (\ref{sol}) we find that the component
fields $\phi^{\alpha_1\ldots \dot\beta_1\ldots}{}^{i_1\ldots i_k
}_{j_1\ldots j_n}(P_\mu)$ in expansion (\ref{sol}) satisfy the
generalized Dirac equations\bl
\begin{equation}\label{Dir-a}
P_{\beta\dot\beta_1}\phi_{\alpha_1\ldots\alpha_k}{}^{\dot\beta_1\dot\beta_2\ldots\dot\beta_n}{}^{i_1\ldots
i_k}_{j_1\ldots j_n}(P_\mu)+ {{\frac{m}{2}}}
\phi_{\beta\alpha_1\ldots\alpha_k}{}^{\dot\beta_2\ldots\dot\beta_n}{}^{i_1\ldots
i_k}_{j_1\ldots j_n}(P_\mu)=0\, ,
\end{equation}
\begin{equation}\label{Dir-b}
P^{\dot\alpha\alpha_1}\phi_{\alpha_1\alpha_2\ldots\alpha_k}{}^{\dot\beta_1\ldots\dot\beta_n}{}^{i_1\ldots
i_k}_{j_1\ldots j_n}(P_\mu)+ {{\frac{m}{2}}}
\phi_{\alpha_2\ldots\alpha_k}{}^{\dot\alpha\dot\beta_1\ldots\dot\beta_n}{}^{i_1\ldots
i_k}_{j_1\ldots j_n}(P_\mu)=0\, ,
\end{equation}\el
as well as the transversality condition
\begin{equation}\label{tr}
P_{\alpha_1\dot\beta_1}\phi^{\alpha_1\ldots\alpha_k
\dot\beta_1\ldots\dot\beta_n}{}^{i_1\ldots i_k}_{j_1\ldots
j_n}(P_\mu)=0\, .
\end{equation}

In order to describe covariant projection of the spin, given by
the eigenvalue equation (\ref{3}), we observe that
\begin{equation}
(\mathcal{D}_3-s_3)(u_{\alpha_1i_1}\cdots
u_{\alpha_ki_k}\overline{u}^{j_1}_{\dbeta_1}\cdots\overline{u}^{j_n}_{\dbeta_n})=0\quad
,
\end{equation}
if
\begin{equation}
s_3=k_1-k_2-(n_1-n_2)\, .
\end{equation}

\section{Examples: $s=\frac{1}{2}$ and $s=1$}\label{Sect.Examples}
\setcounter{equation}{0}

{\it i) Spin $s=1/2$.}

In this case the field (\ref{sol}) is the following\footnote{In this
section all fields $\phi_{\alpha i}$, $\phi^{\dot{\alpha} i}$ etc.
depend on the composite four-momenta $P_{\mu}$ (see (\ref{Pmu})),
what we shall not indicate explicitly.}
\begin{equation}\label{sol-1/2}
\tilde\Phi(u_{\alpha k}, \bar u_{\dot\alpha}^k)= u_{\alpha i}\,
\phi^{\alpha}{}^{i} + \bar u_{\dot\alpha}^{i}\,
\phi^{\dot\alpha}_{i}= -u_{\alpha}^i\, \phi^{\alpha}_{i} + \bar
u^{\dot\alpha}_{i}\, \phi_{\dot\alpha}^{i}\, .
\end{equation}
Inserting in (\ref{sol-1/2})
$u_{\alpha}^i={{\frac{2}{m}}}P_{\alpha\dot\alpha}\bar
u^{\dot\alpha i}$, $\bar
u^{\dot\alpha}_i={{\frac{2}{m}}}P^{\dot\alpha\alpha} u_{\alpha i}$
(see (\ref{D-u})) we obtain
\begin{equation}\label{sol-1/2a}
\tilde\Phi(u_{\alpha k}, \bar u_{\dot\alpha}^k)=
-{{\frac{2}{m}}}P_{\alpha\dot\alpha}\bar u^{\dot\alpha i}\,
\phi^{\alpha}_{i} + {{\frac{2}{m}}}P^{\dot\alpha\alpha} u_{\alpha
i}\, \phi_{\dot\alpha}^{i} = -{{\frac{2}{m}}}\bar u_{\dot\alpha}^{i}
P^{\dot\alpha\alpha}\, \phi_{\alpha i} +
{{\frac{2}{m}}}P^{\dot\alpha\alpha} u_{\alpha i}\,
\phi_{\dot\alpha}^{i}\, .
\end{equation}
{}From (\ref{sol-1/2}) and (\ref{sol-1/2a}) we obtain
$$
u_{\alpha i}( \phi^{\alpha}{}^{i}
-{{\frac{2}{m}}}P^{\dot\alpha\alpha} \, \phi_{\dot\alpha}^{i})+ \bar
u_{\dot\alpha}^{i}\, (\phi^{\dot\alpha}_{i} +{{\frac{2}{m}}}
P^{\dot\alpha\alpha}\, \phi_{\alpha i})=0\, ,
$$
or equivalently
\begin{equation}\label{Dir-1/2}
P_{\alpha\dot\alpha}\phi^{\dot\alpha i}+ {{\frac{m}{2}}}
\phi_{\alpha}^i=0\, , \qquad P^{\dot\alpha\alpha} \phi_{\alpha
i}+ {{\frac{m}{2}}}\phi^{\dot\alpha}_i=0\, .
\end{equation}
Denoting\bl
\begin{equation}
\phi_{\alpha}^1=\phi_{\alpha 2}\equiv\phi_{\alpha}\, , \qquad
\phi_{\alpha}^2=-\phi_{\alpha 1}\equiv\chi_{\alpha}\, ,
\end{equation}
\begin{equation}\label{comp}
\phi_{\dot\alpha 1}=-\phi_{\dot\alpha}^2\equiv\phi_{\dot\alpha}\quad
, \qquad \phi_{\dot\alpha
2}=\phi_{\dot\alpha}^1\equiv\chi_{\dot\alpha}\, ,
\end{equation}\el
($\epsilon^{12}=+1$ in our notation) the equations (\ref{Dir-1/2})
are (we remind that in our paper
$P_{\alpha\dot\alpha}=\frac{1}{2}P_\mu\sigma^\mu_{\alpha\dot\alpha}$)\bl
\begin{equation}\label{Dir-1/2-1}
P_\mu\sigma^\mu_{\alpha\dot\alpha}\chi^{\dot\alpha}+ m
\phi_{\alpha}=0\, , \qquad P_\mu\sigma^{\mu\dot\alpha\alpha}
\phi_{\alpha}+ m\chi^{\dot\alpha}=0\, ,
\end{equation}
\begin{equation}\label{Dir-1/2-2}
P_\mu\sigma^\mu_{\alpha\dot\alpha}\phi^{\dot\alpha}- m
\chi_{\alpha}=0\, , \qquad P_\mu\sigma^{\mu\dot\alpha\alpha}
\chi_{\alpha}- m\phi^{\dot\alpha}=0\, .
\end{equation}\el
If we  define
\begin{equation}\label{real-1}
\phi^{\dot\alpha}_{i}=\bar\phi^{\dot\alpha}_{i}=
\overline{(\phi^{\alpha}{}^{i})}\, ,
\end{equation}
{\it i.e.},
\begin{equation}\label{real-2}
\phi^{\dot\alpha}=\bar\phi^{\dot\alpha}=
\overline{(\phi^{\alpha})}\, ,\qquad
\chi^{\dot\alpha}=\bar\chi^{\dot\alpha}=
\overline{(\chi^{\alpha})}\, ,
\end{equation}
one can pass to four-component Dirac spinors
$(\overline{\psi}=\psi^\dag\gamma_0)$
\begin{equation}\label{Dir-spinor}
\psi_1=\psi\equiv \left(
         \begin{array}{c}
           \phi_\alpha \\
           \chi^{\dot\alpha} \\
         \end{array}
       \right)\, ,\qquad
\psi_2=\psi^C=C\overline{\psi}\equiv \left(
         \begin{array}{c}
           \chi_\alpha \\
           -\phi^{\dot\alpha} \\
         \end{array}
       \right)\, ,
\end{equation}
and use the Dirac matrices $\gamma_\mu$ in Weyl representation
\begin{equation}
\gamma_\mu=\left(
             \begin{array}{cc}
               0 & \sigma^\mu_{\alpha\dot\beta} \\
               \sigma^{\mu{\dot\alpha\beta}} & 0 \\
             \end{array}
           \right)\, , \qquad \{\gamma_\mu,
           \gamma_\nu\}=2\eta_{\mu\nu}\, ,
\end{equation}
where
\begin{equation}
C\gamma_\mu^T=-\gamma_\mu C \Rightarrow C=\gamma_2\gamma_0=\left(
\begin{array}{cc}
  \epsilon^{\alpha\beta} & 0 \\
  0 & -\epsilon_{\dot\alpha\dot\beta} \\
\end{array}
\right)\, ,
\end{equation}
and then, the equations (\ref{Dir-1/2-1}), (\ref{Dir-1/2-2}) take
the form
\begin{equation}\label{Dir-1/2-f}
(P_\mu\gamma^\mu+ m)\, \psi_1=0\, , \qquad (P_\mu\gamma^\mu+ m)
\,\psi_2=0\, ,
\end{equation}
where the field $\psi_1$ describes free relativistic spin
$\frac{1}{2}$ particles, and $\psi_2$ its charge-conjugated
counterpart.

One can note that the two Dirac fields (\ref{Dir-spinor}) form a
$SU(2)$-pseudo-Majorana spinor (these spinors are in $D=1+3$
dimension \cite{KugoT})
\begin{equation}\label{Maj-spinor}
\psi^i\equiv \left(
         \begin{array}{c}
           \phi_\alpha^i \\
           \bar\phi^{\dot\alpha i} \\
         \end{array}
       \right)\, ,
\end{equation}
which satisfies the reality condition
\begin{equation}\label{Maj-cond}
\psi^i{}^T C\gamma_5=\epsilon^{ij}\bar\psi_j\, ,
\end{equation}
where
\begin{equation}
\gamma_5=-i\gamma_0\gamma_1\gamma_2\gamma_3=\left(
             \begin{array}{cc}
               I_2 & 0 \\
               0 & -I_2 \\
             \end{array}
           \right)\, .
\end{equation}
The presence of $\gamma_5$ in (\ref{Maj-cond}) reflects the `pseudo'
reality.

{\it ii) Spin $s=1$.}

In this case the field (\ref{sol}) is
\begin{equation}\label{sol-1}
\tilde\Phi(u_{\alpha k}, \bar u_{\dot\alpha}^k)= {\frac{1}{2}}\,
u_{\alpha i} u_{\beta j} \, \phi^{\alpha\beta}{}^{ij}+ u_{\alpha
i}\bar u_{\dot\beta}^{j}\, \phi^{\alpha \dot\beta}{}^{i}_{j}+
{\frac{1}{2}}\, \bar u_{\dot\alpha}^{i}\bar u_{\dot\beta}^{j}\,
\phi^{\dot\alpha\dot\beta}{}_{ij}\, .
\end{equation}
Inserting in this expression $u_{\alpha i
}={{\frac{2}{m}}}P_{\alpha\dot\alpha}\bar u^{\dot\alpha}_i$, $\bar
u_{\dot\alpha}^i=-{{\frac{2}{m}}}P_{\alpha\dot\alpha} u^{\alpha i}$,
we obtain
\begin{equation}\label{sol-1a}
\tilde\Phi(u_{\alpha k}, \bar u_{\dot\alpha}^k)=
{\frac{1}{2}}\,{\frac{2}{m}}\, \left(u_{\alpha i} u_{\beta j} \,
P^{\dot\beta\beta}\phi^{\alpha}_{\dot\beta}{}^{ij}+ u_{\alpha i}\bar
u_{\dot\beta}^{j}\,
(-P^{\dot\beta\beta}\phi^{\alpha}_{\beta}{}^{i}_{j}
+P^{\dot\alpha\alpha}\phi^{\dot\beta}_{\dot\alpha}{}^{i}_{j} )-
 \bar u_{\dot\alpha}^{i}\bar
u_{\dot\beta}^{j}\,
P^{\dot\alpha\alpha}\phi^{\dot\beta}_{\alpha}{}_{ij}\right)\, .
\end{equation}
Comparing  (\ref{sol-1}) and (\ref{sol-1a}) we
obtain the following equations\bl
\begin{equation}\label{Dir-1a}
P_{\alpha\dot\alpha}\phi^{\dot\alpha}_{\beta}{}^{ij}+
{{\frac{m}{2}}} \phi_{\alpha\beta}{}^{ij}=0\, , \qquad
P^{\dot\alpha\alpha} \phi_{\alpha}^{\dot \beta}{}^{ij}+
{{\frac{m}{2}}}\phi^{\dot\alpha\dot\beta}{}^{ij}=0\, ,
\end{equation}
\begin{equation}\label{Dir-1b}
{\frac{1}{2}}\,(P_{\alpha\dot\alpha}\phi^{\dot\alpha
\dot\beta}{}^{ij}+  P^{\dot\beta\beta} \phi_{\alpha\beta}{}^{ij})+
{{\frac{m}{2}}}\phi_{\alpha}^{\dot\beta}{}^{ij}=0\, .
\end{equation}\el
The antisymmetric parts of equations (\ref{Dir-1a}) provide the
transversality condition for fields $\phi^{\alpha \dot\beta}{}_i^j$
\begin{equation}\label{trans}
P_{\alpha\dot\beta}\phi^{\alpha \dot\beta}{}^{ij}=0\, .
\end{equation}
Using $P_{\alpha\dot\beta}P^{\dot\beta\beta}=
{{\frac{1}{4}}}m^2\delta_\alpha^\beta$ we obtain further
\begin{equation}\label{Dir-1c}
P_{\alpha\dot\alpha}\phi^{\dot\alpha \dot\beta}{}^{ij}+
{{\frac{m}{2}}}\phi_{\alpha}^{\dot\beta}{}^{ij}=0\, , \qquad
P^{\dot\beta\beta} \phi_{\alpha\beta}{}^{ij}+
{{\frac{m}{2}}}\phi_{\alpha}^{\dot\beta}{}^{ij}=0\, .
\end{equation}

The equations (\ref{Dir-1a})-(\ref{Dir-1c}) are Bargman-Wigner
equations written in two-spinor notation. One can pass to
four-component Dirac spinor notation if one constructs from the
fields $\phi_{\alpha\beta}{}^{ij}$,
$\phi^{\dot\alpha\dot\beta}{}^{ij}$,
$\phi_{\alpha}{}^{\dot\beta}{}^{ij}$ and
$\phi^{\dot\beta}{}_{\alpha}{}{}^{ij}\equiv
\phi_{\alpha}{}^{\dot\beta}{}{}^{ij}$ the following Bargman-Wigner
fields
\begin{equation} \label{BarV}
\psi_{ab}{}^{ij}= \left(
\begin{array}{c}
\phi_{\alpha b}{}{}^{ij} \\
\phi^{\dot\alpha}{}_{b}{}^{ij} \\
\end{array}
\right)= \left(
\begin{array}{c}
\phi_{a\beta}{}^{ij} \\
\phi_{a}{}^{\dot\beta}{}^{ij} \\
\end{array}
\right)\, ,
\end{equation}
with double Dirac indices $a,b=1,2,3,4$. Since
$\phi_{\alpha\beta}{}^{ij}=\phi_{\beta\alpha}{}^{ij}$,
$\phi^{\dot\alpha\dot\beta}{}^{ij}=\phi^{\dot\beta\dot\alpha}{}^{ij}$
the fields (\ref{BarV}) are symmetric,
$\psi_{ab}{}{}^{ij}=\psi_{ba}{}^{ij}$. Due to the equations
(\ref{Dir-1a})-(\ref{Dir-1c}) the fields (\ref{BarV}) satisfy the
Bargmann-Wigner-Dirac equation for massive spin 1 fields
$$
P^\mu \,\gamma_{\mu a}{}^{b}
\,\psi_{bc}{}^{ij}+m\psi_{ac}{}^{ij}=0\, .
$$

We obtain Proca fields if we define the fields
\begin{equation} \label{Proca}
A_\mu{}^{ij} =
\sigma_\mu{}^\alpha_{\dot\beta}\phi_\alpha{}^{\dot\beta}{}^{ij}\quad
,\qquad F_{\mu\nu}{}^{ij} = m(\sigma_{\mu\nu}{}^{\alpha}_{\beta}
\phi_{\alpha}^{\beta}{}^{ij}
+\bar\sigma_{\mu\nu}{}^{\dot\alpha}_{\dot\beta}
\phi_{\dot\alpha}^{\dot\beta}{}^{ij})\, .
\end{equation}
Inserting (\ref{Proca}) into the equations
(\ref{Dir-1a})-(\ref{Dir-1c}) we obtain the Proca equations\bl
\begin{equation} \label{Pr1}
P^\mu A_\mu{}^{ij}=0\, ,
\end{equation}
\begin{equation} \label{Pr2}
P_\mu A_\nu{}^{ij} -P_\nu A_\mu{}^{ij} = F_{\mu\nu}{}^{ij}\, ,
\end{equation}
\begin{equation} \label{Pr3}
P^\mu F_{\mu\nu}{}^{ij}-m^2 A_\nu{}^{ij} =0\, ,
\end{equation}\el
as well as the identity
\begin{equation} \label{t}
P_{[\,\mu} F_{\nu\lambda]}{}^{ij} =0\, .
\end{equation}

We obtained three complex fields (internal $SU(2)$-triplet) with
spin $s=1$. On the function (\ref{sol-1}) we should impose  the
reality condition $\widetilde\Phi=\overline{\widetilde\Phi}$ which
gives
\begin{equation}\label{real-1a}
\phi^{\dot\alpha\dot\beta}{}_{ij}=\bar\phi^{\dot\alpha\dot\beta}{}_{ij}=
\overline{(\phi^{\alpha\beta}{}^{ij})}\, , \qquad \phi^{\alpha
\dot\beta}{}^{i}_{j}=\overline{(\phi^{\beta\dot\alpha}{}^{j}_{i})}
\, .
\end{equation}
The relations (\ref{Pr2}) can be written down as the following
$SU(2)$-Majorana reality conditions
\begin{equation}
\psi^{ij}_{cd}(C\gamma_5)_{ca}(C\gamma_5)_{db}=\epsilon^{ik}\epsilon^{jl}\overline{\psi}_{klab}
\end{equation}
and the fields (\ref{Proca}) satisfy the reality conditions
\begin{equation} \label{Proca-r}
\overline{(A_\mu{}^{ij})}=A_\mu{}_{ij}\, ,\qquad
\overline{(F_{\mu\nu}{}^{ij})}=F_{\mu\nu}{}_{ij}\, .
\end{equation}
The relation (\ref{Proca-r}) defines three real vector fields and
the corresponding three real field strengths.

\section{Conclusions}\label{Sect.Conclusions}
\setcounter{equation}{0}

In the present paper we have described a classical and
first-quantized  model of massive relativistic particles with spin
based on a hybrid geometry of phase space, with primary spacetime
coordinates $x_\mu$ and composite four-momenta $P_\mu$ expressed in
terms of fundamental spinorial variables. These spinorial
coordinates describe half of the two-twistor spinorial degrees of
freedom. One can say that the employed geometric framework is half
way between the purely twistorial and the standard spacetime
approaches. We would like to point out  that a model for massive
particles with spin in an enlarged spacetime  derived from
two-twistor geometry, with primary both spacetime coordinates and
four-momenta $P_\mu$, has been recently described in
\cite{BeAzLuMi04}-\cite{AzFrLuMi}. It should be added that the
two-twistor  degrees of freedom were studied recently by one of the
authors  and applied to the spacetime description of massive
spinning particles \cite{FeZi01}-\cite{Fed95}. The difference with
our approach here consists in the choice of the primary geometric
variables which in \cite{FeZi01}-\cite{Fed95} contains, besides
two-twistor degrees of freedom, a primary internal $SU(2)$ spinor,
called the index spinor \cite{FeZi95b}. In the present paper all the
degrees of freedom describing massive particles with spin and
internal charge are derived   entirely from the two-twistor
geometry.

One of the features of the description of spin in the twistor
framework as well as in our model is the use of an orthogonal
reference frame in four-momentum space, with three basic
four-vectors $P_\mu^{(r)}$ orthogonal to the fourth four-vector
$P_\mu$ (see (\ref{5bisnoe})). In such a way the relativistic spin
in any frame is described by the $SU(2)$ algebra of
Lorentz-invariant spin projection operators. Consequently, we
describe the state with definite values of spin square
$\mathbf{s}^2$ and  invariant spin projection $s_3$  by a
Lorentz-covariant wave function.

In order to quantize the classical system we have introduced a
complete set of commuting observables, which determine the
generalized coordinates of the wavefunction. In our case the set of
commuting generalized coordinates does not contain all the spacetime
coordinates, because in  our geometric framework they do not commute
(see (\ref{X.Non-commutative})). As a result, only the
Lorentz-invariant projection $\widetilde{x}_0=x_\mu P^\mu$ can be
included into the quantum-mechanical commuting coordinates. In such
a way we are allowed to use plane waves $e^{ix_\mu P^\mu}$ as
describing the spacetime dependence of the wavefunction. We
conclude, therefore, that although in our framework the spacetime
coordinates of spinning massive particles are non-commutative, we
are able to obtain the standard plane wave solutions.

\bigskip

\subsubsection*{Acknowledgements}
The authors wish to thank J.A. de Azc\'{a}rraga for reading the
manuscript and many valuable comments. Two of the authors (A.F. and
J.L.) would like to acknowledge the financial support of KBN grant 1
P03B 01828.  This work has also been supported by the Spanish
Ministerio de Educaci\'{o}n y  Ciencia through grant FIS2005-02761,
EU FEDER funds, the Generalitat Valenciana and the Poland-Spain
scientific cooperation agreement. One of us (C.M.E.) wishes to thank
the Spanish M.E.C. for his research grant.

\begin{appendix}
\section{Appendix: Time evolution of the constraints}\label{Appendix}
The equations describing the time evolution of all the constraints
are\bl
\begin{eqnarray}
\dot D_r = \{ {\cal H}, D_r\}& = & \epsilon_{rpq}\lambda^{(D)}_p
(D_q-s_q) + \epsilon_{rpq}\lambda^{(B)}_p B_q +
\epsilon_{rpq}\lambda^{(C)}_p C_q
- {\frac{1}{2}}\lambda^{(R)}_r \approx \nonumber \\
&\approx&-\epsilon_{rpq}\lambda^{(D)}_p s_q - {\frac{1}{2}}
\lambda^{(R)}_r=0\, , \label{HD}\\
\dot D_0 = \{ {\cal H}, D_0\}& = &
- {\frac{1}{2}}\lambda^{(R)}_0=0\, , \label{HD0}\\
\dot B_r = \{ {\cal H}, B_r\}& = & \epsilon_{rpq}\lambda^{(B)}_p
(D_q-s_q) + \epsilon_{rpq}\lambda^{(D)}_p B_q - i \lambda^{(C)}_r
(m^2-C_0)
+ \lambda^{(C)}_0 C_r \approx \nonumber \\
&\approx&-\epsilon_{rpq}\lambda^{(B)}_p s_q - i m^2
\lambda^{(C)}_r =0\, , \label{HB}\\
\dot B_0 = \{ {\cal H}, B_0\}& = & i \lambda^{(C)}_r C_r - i
\lambda^{(C)}_0 (m^2-C_0)
+2i \lambda^{(T)}(m^2+T)\approx \nonumber \\
&\approx& - i m^2 \lambda^{(C)}_0 +2i m^2\lambda^{(T)} =0\, ,
\label{HB0}\\
\dot C_r = \{ {\cal H}, C_r\}& = & \epsilon_{rpq}\lambda^{(D)}_p C_q
+ i \lambda^{(B)}_r (m^2-C_0) -i \lambda^{(B)}_0 C_r \approx
i m^2 \lambda^{(B)}_r =0\, ,\nonumber\\&& \label{HC}\\
\dot C_0 = \{ {\cal H}, C_0\}& = & -i \lambda^{(B)}_r C_r + i
\lambda^{(B)}_0 (m^2-C_0) \approx i m^2 \lambda^{(B)}_0 =0\, ,
\label{HC0}\\
\dot R_r = \{ {\cal H}, R_r\}& = & {\frac{1}{2}}\lambda^{(D)}_r +
\lambda^{(S)}s_r +{\frac{1}{2}}\delta_{r3}
\lambda^{(S_3)} =0\, , \label{HR}\\
\dot R_0 = \{ {\cal H}, R_0\}& = & {\frac{1}{2}}\lambda^{(D)}_0 +
{\frac{1}{2}}\lambda^{(Q)}
=0\, , \label{HR0}\\
\dot T = \{ {\cal H}, T\}& = & -2 i \lambda^{(B)}_0 (m^2+T) \approx
-2 i m^2 \lambda^{(B)}_0 =0\, , \label{HT}\\
\dot S = \{ {\cal H}, S\}& = & - \lambda^{(R)}_r s_r =0\, ,
\label{HS}\\
\dot S_3 = \{ {\cal H}, S_3\}& = &
-{\frac{1}{2}}\lambda^{(R)}_3 =0\, , \label{HS3}\\
\dot Q = \{ {\cal H}, Q\}& = & -{\frac{1}{2}}\lambda^{(R)}_0 =0\quad
. \label{HQ}
\end{eqnarray}\el

One obtains from (\ref{HD})-(\ref{HQ}) the relations which imply the
preservation of the constraints in time:\bl
\begin{eqnarray}
\lambda_0^{(R)}=\lambda_r^{(B)}=\lambda_0^{(B)}=\lambda_r^{(C)}&=&0\quad,\\
\lambda_3^{(R)}&=&0\quad,\label{lambda3R}
\end{eqnarray}\el
\bl
\begin{eqnarray}
\lambda_0^{(C)}&=&2\lambda^{(T)}\, ,\\
\lambda_0^{(Q)}&=&-\lambda_0^{(D)}\, ,\\
\lambda_r^{(R)}&=&-2\epsilon_{rst}\lambda_s^{(D)}s_t\, ,\label{lambdarR}\\
\lambda_r^{(D)}&=&-2\lambda^{(S)}s_r-\delta_{3r}\lambda^{(S_3)}\quad
.\label{lambdarD}
\end{eqnarray}\el
{}From (\ref{lambdarR})-(\ref{lambdarD}) one obtains
\begin{equation}
\lambda_r^{(R)}=2\epsilon_{r3t}\lambda^{(S_3)}s_t\, ,
\end{equation}
in consistency with the relation (\ref{lambda3R}).
\end{appendix}

\end{document}